\documentclass[a4paper,12pt,english]{article}
\pdfoutput=1

\usepackage[bindingoffset=0.3cm,textheight=22.5cm,hdivide={2.7cm,*,2.7cm}, vdivide={*,22cm,*}]{geometry}
\usepackage[bookmarksnumbered=true,breaklinks=true]{hyperref}

\usepackage{amsmath,amsfonts,amssymb,babel,slashed,graphicx,color}

\usepackage[numbers,square,comma,sort&compress]{natbib}

\makeatletter

\renewcommand{\Re}{\mathop{\text{Re}}}

\newcommand{\bbm}{\left(\begin{matrix}}
\newcommand{\ebm}{\end{matrix}\right)}
\newcommand{\beq}{\begin{eqnarray}}
\newcommand{\eeq}{\end{eqnarray}}
\makeatother

\newcommand{\be}{\begin{equation}}
\newcommand{\ee}{\end{equation}}

\newcommand{\beqa}{\begin{eqnarray}}
\newcommand{\eeqa}{\end{eqnarray}} \newcommand{\eq}[1]{(\ref{#1})}
\def\nn{\nonumber} \def \bea{\begin{eqnarray}} \def\eea{\end{eqnarray}}
\def\obar{\overline}
\newcommand{\barr}{\begin{array}}
\newcommand{\earr}{\end{array}}
\numberwithin{equation}{section}

\def\a{\alpha}  \def\b{\beta}
 \def\g{\gamma} \def\G{\Gamma}
 \def\d{\delta} 

\def\l{\lambda} \def\L{\Lambda}  
\def\n{\nu}

\def\cA{{\cal A}}  \def\cC{{\cal C}} 
   
 \def\cH{{\cal H}}  
   
\def\cM{{\cal M}} \def\cN{{\cal N}}  
   
  \def\cU{{\cal U}}

\def\R{{\mathbb R}} \def\C{{\mathbb C}} \def\N{{\mathbb N}}

 \def\one{\mbox{1 \kern-.59em {\rm l}}}

\def\mmu{\mathfrak{u}}

\def\msu{\mathfrak{su}}
\def\mso{\mathfrak{so}}

\def\bit{\begin{itemize}} \def\eit{\end{itemize}} \def\Tr{\mbox{Tr}}

\def\({\left(} \def\){\right)}    

\def\und{\underline}

 \allowdisplaybreaks[3]

\begin{document}

\makeatother


\parindent=0cm

\renewcommand{\title}[1]{\vspace{10mm}\noindent{\Large{\bf

#1}}\vspace{8mm}} \newcommand{\authors}[1]{\noindent{\large

#1}\vspace{5mm}} \newcommand{\address}[1]{{\itshape #1\vspace{2mm}}}


\begin{titlepage}
\begin{flushright}
 UWThPh-2015-28 
\end{flushright}
\begin{center}
\title{ {\Large One-loop stabilization of the fuzzy four-sphere \\[1ex] via softly broken SUSY} }

\vskip 3mm

\authors{Harold C. Steinacker{\footnote{harold.steinacker@univie.ac.at}}
}
 
\vskip 3mm

 \address{ 

{\it Faculty of Physics, University of Vienna\\
Boltzmanngasse 5, A-1090 Vienna, Austria  }  
  }

\bigskip

\vskip 1.4cm

\textbf{Abstract}
\vskip 3mm

\begin{minipage}{14cm}%

We describe a stabilization mechanism for fuzzy $S^4_N$ in the Euclidean IIB matrix model 
due to vacuum energy in the 
presence of a {\em positive} mass term.  The one-loop effective potential for the radius contains 
an attractive contribution attributed to supergravity, while the 
mass term induces a repulsive contribution for small radius due to SUSY breaking.
This leads to a stabilization of the radius. The mechanism 
should be pertinent to  recent 
results on the genesis of 3+1-dimensional space-time in the Minkowskian IIB model.

\end{minipage}

\end{center}

\end{titlepage}

\tableofcontents

\section{Introduction}

It is widely expected that space-time should have some kind of quantum structure at short distances. 
On the other hand, we known that (local) Lorentz invariance is respected to a very high precision. Reconciling these 
two requirements is a non-trivial task.
In two dimensions,  the fuzzy sphere  $S^2_N$ \cite{Madore:1991bw,hoppe}
or fuzzy (anti-) de Sitter space  $(A)dS^2_N$ \cite{Ho:2000br,Jurman:2013ota}
provide examples of quantum geometries which are compatible with the full isometry group 
$SO(3)$ and $SO(2,1)$, respectively. This is possible because the underlying Poisson structures are 
invariant.  However in 4 dimensions, any Poisson structure will necessarily break the local isometry group;
in fact, $S^4$ does not even admit any symplectic structure at all.

Nevertheless, there is a four-dimensional version of the fuzzy sphere $S^4_N$ 
\cite{Grosse:1996mz,Castelino:1997rv} which  is fully covariant under $SO(5)$, 
and has a finite number of ``quantum cells''.
The price to pay is a tower of  higher spin modes \cite{Ramgoolam:2001zx}, which arise because
$S^4_N$ has internal structure and is more properly described as a $S^2_{N+1}$ bundle over $S^4$ 
\cite{Ho:2001as,Kimura:2002nq,Medina:2002pc,Abe:2004sa,Valtancoli:2002sm}, or 
as a twisted stack of $N+1$ four-spheres. Nevertheless
 $S^4_N$ can be realized in matrix models, and one may hope that these 
higher-spin modes acquire a mass via quantum effects in that setting.
If that turns out to be true, this type of background may allow to reconcile  the
ideas of emergent gravity in matrix models \cite{Steinacker:2010rh,Steinacker:2012ra}
with Lorentz invariance. 
It has indeed been argued on different grounds that gravity should emerge on a Lorentzian analog of this space   \cite{Heckman:2014xha}.

In the present paper, we consider the realization of fuzzy $S^4_N$ within the IIB or 
IKKT matrix model \cite{Ishibashi:1996xs}, taking into account quantum corrections.
This is motivated in part by the remarkable recent results  of computer-simulation of the 
Lorentzian IIB model. It was found \cite{Kim:2011cr} that a 3+1-dimensional extended geometry arises dynamically in this model, 
interpreted as emergent cosmological space-time. To define the Lorentzian model, an IR regularization of the 
model is required. Such a regularization can be viewed as analytic continuation of a bosonic mass term\footnote{This regularization 
might also implement Feynman's $i\epsilon$ prescription in the matrix model.}.
This suggests that to understand these findings and the relation with the Euclidean case, 
an analogous mass term should also be introduced in the Euclidean IIB model.

The main result of the present paper is that introducing a {\em positive mass} term in the Euclidean IIB model
leads indeed to an interesting stabilization mechanism for  fuzzy geometries at one loop, which in particular 
stabilizes the radius of $S^4_N$. This is quite remarkable and perhaps surprising: 
at the classical level $S^4_N$ is certainly not a solution, since the radius is not stabilized.
It would formally be a solution for 
a negative mass term, but then the model becomes unstable, so this is not an option.
One might try to stabilize it by adding a ``flux'' term of order 5 to the model \cite{Kimura:2002nq}, 
however this spoils the good UV properties of the model, and stabilization is lost at the quantum level \cite{Azuma:2004yg}. 

Remarkably, adding a positive mass term to the {\em supersymmetric} IIB model
does provide a stabilization mechanism at one-loop: Without mass term, 
the residual (finite) one-loop effects lead to an attractive (negative) contribution to the effective potential,
which may be interpreted in terms of supergravity \cite{Ishibashi:1996xs,Chepelev:1997av}.
However this effect is independent of the radius, and does not stabilize $S^4_N$ in the IIB model.
Upon adding a (small) positive mass term,
the bosonic quantum fluctuations are suppressed over the fermionic fluctuations, so that the latter 
lead to a  positive contribution to the  effective potential. This effect is large 
for small radius $r$ of $S^4_N$ as the SUSY breaking is significant, but it is small for large $r$
since then the SUSY breaking is insignificant, and the attractive contribution attributed to supergravity prevails.
As a result, the effective potential has a stable minimum as a function of $r$, thus 
stabilizing  $S^4_N$. 
We elaborate this effect in section \ref{sec:one-loop}
by performing a detailed one-loop computation, integrating out all fluctuation 
modes except for the lowest (would-be)  massless modes. The result is argued to be robust for large $N$,
and it only applies to the maximally supersymmetric IIB model.

The mechanism under consideration is not restricted to the particular $S^4_N$ background. 
However, it is very plausible that the $S^4_N$ should be preferred dynamically
over less symmetric spaces such as $T^2\times T^2$ \cite{Bal:2004ai} 
or $S^2\times S^2$ or $\C P^2$. Moreover,
the stabilization mechanism does not apply to  branes with dimension 6 or higher. 
We also find in section \ref{sec:fuzzyS2} that for suitable parameters,
fuzzy $S^4_N$ has lower energy than a fuzzy two-sphere.
On the other hand, we also find 
 a preference towards high-rank gauge groups i.e. stacks of multiple branes. 
Thus we certainly cannot draw any non-perturbative conclusions. Nevertheless, the observed
stabilization of $S^4_N$ at one loop indicates that this semi-classical geometry may be a good step towards
unraveling the structure of space-time.

\section{Fuzzy $S^4_N$}

Let $\g_i, \ i=1,...,5$ be the gamma matrices associated to $SO(5)$, which act on $\C^4$. 
Fuzzy $S^4_N$ \cite{Grosse:1996mz,Castelino:1997rv} is defined in terms of the following ``many-particle'' 
version\footnote{We include a factor $\frac 12$ to avoid excessive factors 4 in subsequent computations.} of $\g_i$ 
\begin{align} 
 X_i &= \frac 12\big(\g_i \otimes 1 \otimes ... \otimes 1 + ... + 1 \otimes ... \otimes 1  \otimes  \g_i\Big) , \qquad i=1,2,...,5 
\end{align}
acting on 
\begin{align}
 \cH_N = \big(\C^4 \otimes ... \otimes \C^4\big)_{\rm sym} =(0,N)_{SO(5)} \ .
\end{align}
We denote  irreducible highest weight representation by the Dynkin indices.
This  can be written succinctly as
\begin{align}
 X_i = \frac 12 a^\dagger_\a (\g_i)^\a_{\ \b} a^\b
 \label{JS-S4}
\end{align}
with 4 bosonic oscillators 
\begin{align}
 [a^\b,a^\dagger_\a] = \d_\a^\b
\end{align}
acting on the Fock space $\cH_N = a^\dagger_{i_1} ... a^\dagger_{i_N}|0\rangle$.
Their commutation relations are obtained as
\begin{align}
 [X_i,X_j] &=: i\, \cM_{ij}  \label{X-i-CR} \\
    [\cM_{ij},\cM_{kl}] &=i(\d_{ik}\cM_{jl} - \d_{il}\cM_{jk} - \d_{jk}\cM_{il} + \d_{jl}\cM_{ik}) 
 \label{M-M-relations} \\
  [\cM_{ij},X_k] &= i(\d_{ik} X_j - \d_{jk} X_i) .
 \label{M-X-relations}
\end{align}
In particular,  
\begin{align}
\cM_{ij} =  a^\dagger M_{ij} a , \qquad 
  M_{ij} = \frac 1{4i}[\g_i,\g_j]
\end{align}
generate the $SO(5)$ rotations on the algebra $End(\cH_N)$ generated by the $X_i$.
This means that fuzzy $S^4_N$ is a Snyder-type noncommutative space \cite{Snyder:1946qz}, where 
$\cM_{ij}$ generate the $SO(5)$ transformations of 
the $X_i$, which transform in the $(1,0)$ of $SO(5)$.

Now consider the $SO(5)$-invariant operator $\sum_i X_i^2$. 
Since $(0,N)$ is irreducible,
it must be $\sim \one$, and the constant is easily found to be \cite{Castelino:1997rv}
\begin{align}
\hat R^2 := \sum_i X_i^2  = \frac 14N(N+4) \one =: r_N^2 \one \ .
 \label{radius}
\end{align}
Together with the $SO(5)$ covariance, this strongly suggests an interpretation in terms of $S^4$
with radius $r_N=\frac 12\sqrt{N(N+4)}$.
Indeed the above construction is entirely analogous to that of the fuzzy sphere \cite{Madore:1991bw},
using $SO(5)$ instead of $SO(3)$.

To gain some intuition, consider the north pole $p\in S^4$, which is invariant under the 
local stabilizer group $SO(4)\subset SO(5)$. We can then decompose the $\cM_{ij}$ into rotation generators
\begin{align}
 \cM_{\mu\nu}, \qquad \mu,\nu\in\{1,2,3,4\}
\end{align}
which generate the ``local'' $SO(4)$, and 
\begin{align}
 P_\mu &= \frac 1{r_N} \cM_{\mu 5} \ \  , \qquad \mu\in\{1,2,3,4\} 
\end{align}
which correspond to ``translations'' moving $p$ along $S^4$; the latter transform as 
vectors of $SO(4) \cong SU(2)_L \times SU(2)_R$.
In particular, \eq{M-X-relations} implies 
\begin{align}
[P_\mu,X_\nu] 
= i\d_{\mu\nu} \frac 1{r_N} X_5 
 \label{p-X-relations}
\end{align}
which reduces to the canonical commutation relations upon replacing $X^5 \to r_N$ at or near $p$, 
and  \eq{M-M-relations} implies 
\begin{align}
[P_\mu,P_\nu] 
 &= -i \frac 1{r_N^2} \cM_{\mu\nu} \ .
\end{align}
Thus the Poincare group
is recovered as usual from $SO(5)$ via an In\"on\"u-Wigner contraction.

\paragraph{Fuzzy $S^4_N$ versus fuzzy $\C P^3_N$.}


Although the above results seem very straightforward, it turns out that fuzzy $S^4_N$
is more properly understood as a degenerate ``projection'' of fuzzy\footnote{There is another, 
similar interpretation of fuzzy $S^4_N$ as deformation of  
$SO(5)/U(2)$ \cite{Ho:2001as},
which also allows to resolve the fuzzy $S^2_N$ fibration over $S^4$.} $\C P^3_N$ 
\cite{Medina:2002pc,Fabinger:2002bk,Abe:2004sa,Valtancoli:2002sm,Karabali:2006eg,Medina:2012cs}. 
This can be seen simply by extending \eq{JS-S4} to the full set of 
$\msu(4)\cong \mso(6)\supset \mso(5)$ generators 
\begin{align}
 X_A = a^\dagger_\a (M_A)^\a_{\ \b} a^\b
 \label{JS-S4-full}
\end{align}
acting on 
\begin{align}
 \cH_N  =(0,N)_{SO(5)} = (0,0,N)_{SU(4)}
\end{align}
where $M_A$ are the $\msu(4)$ generators acting on $\C^4$.
Organizing $A = (a,b)_{a<b\leq 6}$ as a basis of $\mso(6)$ and identifying
\begin{align}
 \g_i = 2 M_{i6} \ ,
 \label{gamma-sigma}
\end{align}
we recover the previous generators $X_i$ \eq{JS-S4} from this extended set of generators.
Moreover, since $\cH_N = (0,N)$ of $SO(5)$ is equivalent to $(0,0,N)$ of $SU(4)$,
\eq{JS-S4-full} can be rewritten as
\begin{align}
 X_A = \pi_N(M_A)
\end{align}
acting on $\cH_N$. This is precisely the  construction of 
fuzzy $\C P^3_N$ \cite{Balachandran:2001dd,CarowWatamura:2004ct}, 
which is a quantization of $\C P^3$ viewed as coadjoint orbit in $\msu(4) \cong \R^{15}$
with Poisson structure 
 $\{X_A,X_B\} = f_{AB}^C X_C$
corresponding to the Kirillov-Kostant symplectic form, with commutation relations
\begin{align}
 [X_A,X_B] = if_{AB}^C X_C \ .
\end{align}
In particular, we recognize the $\cM_{ij}$  \eq{X-i-CR} as generators of 
$\mso(5) \subset \mso(6) = \msu(4)$ acting on $\cH_N$.
More importantly, the original $X_i,\ i=1,...,5$ are now recognized as a subset of the 
15 quantized embedding functions 
\begin{align}
 X_A \sim x_A: \quad \C P^3 \hookrightarrow \msu(4) \cong \R^{15}
\end{align}
of the 6-dimensional coadjoint orbit $\C P^3$ in $\R^{15}$.
Here $\sim$ indicates the semi-classical limit.
Dropping the extra generators amounts to a projection $\C P^3\subset\R^{15} \to S^4\subset \R^5$. 
This corresponds to the Hopf map $\C P^3 \to S^4$ along a $S^2$ fiber, as 
discussed in appendix \ref{sec:appendix-Hopf}.

The bottom line is that fuzzy $S^4_N$ is {\em not} the quantization of $S^4$ with some
Poisson structure,
but it is a 4-dimensional degenerate embedding of fuzzy $\C P^3_N$  with squashed 
$S^2_{N+1}$ fiber.
The ``local noncommutativity'' $\cM_{\mu\nu}$ can then be understood as a fibration with 
self-dual 2-forms on $S^4$ which vary\footnote{Such a ``continuous superposition'' of 
noncommutativity was already contemplated in  
\cite{Doplicher:1994tu}.
The present  realization should help to understand better the mechanism 
how physics and in particular gravity can
emerge on such a noncommutative space within matrix models \cite{Steinacker:2010rh,Heckman:2014xha}.} 
along the $S^2$ fiber, as explained in more detail below. 
This allows to preserve full covariance under $SO(5)$, and explains its
somewhat  complicated algebraic structure including apparent non-associativity 
\cite{Ramgoolam:2001zx,Ho:2001as} and higher-spin fields.

\paragraph{Useful results on the fuzzy $S^4$ algebra.}

We collect some  more results for fuzzy $S^4_N$. From
\eq{M-X-relations} one obtains immediately
\begin{align}
 \Box_X X_i \equiv [X_j,[X_j,X_i]] 
  = \, 4 \, X_i   \ .
   \label{box-x}
\end{align}
We also note the  identity
\begin{align}
 \cM_{ij} X_j + X_j \cM_{ij} = [X_i, r_N^2] = 0
 \label{M-X-contract}
\end{align}
which states that the $\cM_{ij}$ are tangential.
Now consider the following Casimir operator
\begin{align}
 \epsilon^{ijklm} X_iX_jX_kX_l X_m  = -\frac 14\cM_{ij}\cM_{kl} X_m \epsilon^{ijklm} =  C^5 \one \ .
\end{align}
To evaluate this, we recall $X_m = \cM_{m6}$  \eq{gamma-sigma}, so that 
\begin{align}
 \cM_{ij}\cM_{kl} X_m \epsilon^{ijklm} 
  &= \frac 16 \cM_{ij}\cM_{kl}\cM_{mn} \epsilon^{ijklmn}  \ .
\end{align}
This is a totally symmetric cubic invariant of $\msu(4)$
which can only be the $d^{A B C} \cM_A \cM_B \cM_C$ tensor,
which for $\C P^3_N$ is given by  \cite{Balachandran:2001dd,CarowWatamura:2004ct}
\begin{align}
 d \cM \cM \cM 
  = \frac{3}{16}(N+2)N(N+4) \ .
\end{align}
Therefore $C^5 = c (N+2) r_N^2$
where $c$ is a constant independent of $N$. This is found to be $c=\frac 14$ for $N=1$,
and it follows that 
\begin{align}
  \epsilon^{ijklm} X_iX_jX_kX_l X_m &= (N+2) r_N^2  
   \label{epsXXXXX}
\end{align}
consistent with \cite{Castelino:1997rv,Kimura:2002nq}.

\subsection{Alternative interpretation: stack of $N+1$ branes}
\label{sec:brane-stack}

We have just explained the interpretation of fuzzy $S^4_N$ as 
``squashed'' fuzzy $ \C P^3_N$ with degenerate fiber $S^2_{N+1}$.
There is an alternative interpretation as a twisted stack of $N+1$ 
spherical branes (cf. \cite{Castelino:1997rv,Ho:2001as,Karczmarek:2015gda}),
carrying a  $\mmu(N+1)$-valued noncommutative Yang-Mills gauge field with flux $\cM^{\mu\nu}$. 
This gauge field configuration is necessarily non-trivial,
since  $S^4$  does not admit any symplectic form due to $H^2(S^4) = 0$.
 As explained in appendix \ref{sec:appendix-Hopf}, the generators $\cM^{\mu\nu}$ of the 
local $\mso(4) = \msu(2)_L \times \msu(2)_R$ act on the local $S^2$ fiber only via
$\msu(2)_L$, while $\msu(2)_R$ acts trivially. 
These $\msu(2)_L$ generators $J_k^{(L)}$ generate the local $S^2_{N+1}$. 
This means that $\cM^{\mu\nu}$ is a self-dual  $\mmu(N+1)$ ``instanton'' configuration
\begin{align} 
\cM^{\mu\nu} &= \epsilon^{\mu\nu k}_L\, J_k^{(L)} \ . 
\end{align}
Another way to see this\footnote{I would like to thank J. Karczmarek for useful discussions on this aspect.}  
is via coherent states \cite{Karczmarek:2015gda}.
As usual, such a $\mmu(N+1)$-valued flux can be interpreted as a gauge field  on a stack of $N+1$ noncommutative 
branes, which are identified e.g. by diagonalizing $J_3^{(L)}$, or by choosing $N+1$ 
coherent states $|i\rangle$ on $S^2_N$.
Then the  modes $|i\rangle\langle j| \in \mmu(N+1)$ can  be interpreted as
``string'' connecting two such sheets, and
we can assign a Poisson structure 
\begin{align}
\langle i|\cM^{\mu\nu}|i\rangle  &\sim  \theta^{\mu\nu} = \{x^\mu,x^\nu\}
\end{align}
to these $N+1$ leaves covering $p\in S^4$.

We can provide a non-trivial consistency check of this semi-classical picture.
To this end, we 
evaluate \eq{epsXXXXX} at the north pole (or via $\langle i| . |i\rangle$) in terms of the Pfaffian   
\begin{align}
 {\rm Pf} \theta^{\mu\nu} = \frac 12\epsilon_{\mu\nu\rho\sigma}\theta^{\mu\nu}\theta^{\rho \sigma}
\end{align}
which gives
 \begin{align}
 (N+2) r_N^2 = 
  \epsilon^{ijkl5} X_iX_jX_kX_l X_5 \sim  r_N \epsilon_{\mu\nu\rho\sigma}\theta^{\mu\nu}\theta^{\rho \sigma}
  = 2 r_N\,  {\rm Pf} \theta^{\mu\nu} \ .
 \end{align}
Thus
 \begin{align}
  {\rm Pf} \theta^{\mu\nu} \  \sim  \  r_N^2 
  \label{Pfaff-estimate}
 \end{align}
 for large $N$.
Now the  semi-classical formula 
$tr \sim  \frac 1{(2\pi)^2} \int \frac 12 \omega^2$ for 4-dimensional symplectic spaces 
applied to the $N+1$ (twisted) leaves gives
 \begin{align} 
 \frac 8{3} r_N^6 &\sim tr  XXXXX \epsilon 
 \sim \frac {N+1}{(2\pi)^2} \int \frac 12 \omega^2 2 r_N\, {\rm Pf}\theta^{\mu\nu}  \nn\\ 
 &=  2 r_N (N+1)\frac 1{(2\pi)^2} Vol(S^4)
 =   \frac{4}{3} r_N^5 (N+1)
\end{align}
 noting that the Pfaffian and $\frac 12\omega^2$ cancel each other, giving
$\int \frac 12 \omega^2 {\rm Pf}\theta^{\mu\nu}=  Vol(S^4) = \frac{8\pi^2}{3}r_N^4$. 
This is consistent  with $r_N\sim \frac N2$  \eq{radius} for large $N$,
confirming the above semi-classical  picture.
In other words, fuzzy $S^4_N$ has approximately $\dim \cH_N \sim \frac{1}6 N^3$
``noncommutative'' volume quanta $L_{NC}^4$ distributed over  $(N+1)$ times the metric volume of $S^4$,
\begin{align}
 \frac 16 N^3 L_{NC}^4 &\sim (N+1)   \frac{8\pi^2}{3}r_N^4 \sim \frac{\pi^2}{6} N^5  \ .
 \end{align}
 Hence the scale of noncommutativity 
 i.e. the effective uncertainty scale 
 is set by\footnote{For the fuzzy sphere $S^2_N$ one finds similarly $L_{NC} \sim \sqrt{N\pi}$, 
 cf. section \ref{sec:fuzzyS2}.} 
 \begin{align}
  L_{NC} &\sim \sqrt{N\pi} \ .
  \label{planck-scale}
\end{align}
On the other hand,  each volume $L_{NC}^4$ in target space is covered by $(N+1)$ such cells,
which reflects the internal structure of $S^4_N$. 
Thus one might also argue that the ``atoms'' of space have the size 
\begin{align}
 L_0^4 &= \frac{16\pi^2}{N^4} r_N^4 , \qquad 
  L_0 \sim \sqrt{\pi} 
  \label{atom-scale}
\end{align}
which is of order one. The physical significance of these two scales and their 
possible relation to the Planck scale remains to be elucidated.

\subsection{Functions and harmonics on fuzzy $S^4$}

The algebra of functions $End(\cH_N)$ decomposes into the $SU(4)$ harmonics
\begin{align}
 End(\cH_N) &= (0,0,N)\otimes (N,0,0) = \bigoplus\limits_{n=0}^N (n,0,n)
\end{align}
which is a truncation of the classical algebra of (polynomial) functions on  $\C P^3$,
\begin{align}
 Pol(\C P^3 ) = \bigoplus\limits_{n=0}^\infty (n,0,n) \ . 
\end{align}
Each of these  $(n,0,n)$ decompose into the $SO(5)$ 
harmonics as follows 
\begin{align}
 (n,0,n) =\bigoplus\limits_{m=0}^n (n-m,2m) \ .
 \label{n0n-decomp}
\end{align}
The $(n,0)$ modes correspond to the (totally symmetrized traceless) 
polynomial functions $P_n(X_i)$ on $S^4$ of degree $n$. 
The $(n-m,2m)$ modes with $m\neq 0$ have 
a non-trivial dependence along the $S^2$ fiber, i.e. they transform non-trivially under the local stabilizer 
group $SU(2)_L\times SU(2)_R$ at fixed $p\in S^4$, hence they correspond to  higher spin modes. 
Thus all ``bosonic'' higher spin modes of $SO(5)$ arise precisely once\footnote{This was first found in 
\cite{Ramgoolam:2001zx} using the language of Young diagrams.} 
in $End(\cH_N)$, while 
the fermionic $(n,k)$ modes with odd $k$ modes will be recovered in section \ref{sec:fermions}.

We would like to have a more explicit description of these modes. 
To this end, we denote the highest weight vector in $(1,0)$ by $X_1^+$, 
and observe that the  $X_5$ generator is stabilized by $SO(4)\subset SO(5)$ 
and corresponds to the weight 0 vector in $(1,0)$. 
Thus the highest weight vector in $(n,0)$ is given by $(X_1^+)^n$.
Clearly this is also the highest weight vector in $(n,0,n)$,
so that all modes in \eq{n0n-decomp} are obtained by acting with 
the universal enveloping algebra $\cU(\msu(4))$ on  $(X_1^+)^n$. 

Now denote with $\cM_L^+$ the highest weight element in 
$(0,2) \subset (1,0,1)$; 
 this is the raising operator of $SU(2)_L\subset SO(4)$.
Clearly $\cM_L^+$ is obtained by acting with 
some $SU(4)$ generator $[X_2^+,.]$ (say) on $X_1^+$.
Similarly, acting with $[X_2^+,.]$ on $(X_1^+)^n \in (n,0,n)$ gives $(X_1^+)^{n-1} \cM_L^+$, 
which  is the highest weight vector of $(n-1,2)$. 
More generally, acting with suitable elements in $\cU(\msu(4))$ on $(X_1^+)^n$  generates 
all the $(X_1^+)^{n-m} (\cM_L^{+})^m$, which are the highest weight vectors in $(n-m,2m)$. 
Thus we have  obtained an explicit description of these higher spin modes:
\begin{align}
 (n-m,2m) = {\rm span}\{ P_{n-m}(X^i)P_m(\cM^{ij}) \} \ .
\end{align}
We can decompose these modes further w.r.t. the ``local'' $SO(4)$ structure e.g. at the north pole
$p\in S^4$, where  $\cM_{ij}$  decomposes into 
the $\cM_{\mu\nu}$ and the $P_\mu$  generators.
Then e.g. the  $(n,2)$ modes on $S^4$
comprise  modes of the form $f(X)P_\mu$  and 
modes of the form $f(X)\cM_{\mu\nu}$. 
Similarly the $(n,4)$ modes comprise e.g. $f(X)P^\mu P^\nu$ fields, etc.

\paragraph{Integration over the fiber.}

The decomposition \eq{n0n-decomp} in particular provides a map which maps any function on $\C P^3_N$ to a spin 0 function on $S^4_N$,
defined by
\begin{align}
 \langle . \rangle:\quad  End(\cH_k) &\to End(\cH_k) \\
 (n,m) &\mapsto (n,0) \ .
\end{align}
Classically, this amounts to integrating over the local fiber $S^2$. 
Using this projection, the space $\oplus (n,0)$ becomes a commutative but non-associative algebra, which 
for $N\to \infty$ becomes the 
classical algebra of functions on $S^4$.
This algebra is discussed in detail in \cite{Ramgoolam:2001zx}.

\subsection{Matrix Laplacian and quadratic Casimirs.}

The eigenvalues of the matrix Laplacian  
\begin{align}
 f \to \Box f =  [X_i,[X_i,f]]
\end{align}
can be obtained as follows: 
recalling the identification $X_i = \cM_{i6}$ and $[X_i,.] = \cM_{i6}^{(ad)}$ on $End(\cH_N)$, we have 
\begin{align}
 \Box  = [X_i,[X_i,.]] = \sum_{i} \cM_{i6}^{(ad)} \cM_{i6}^{(ad)}  = C^2[\mso(6)]^{(ad)} - C^2[\mso(5)]^{(ad)}
 \label{box-Casimirs}
\end{align}
in terms of the quadratic Casimirs
\begin{align}
 C^2[\mso(6)] &= \sum_{a<b\leq 6} \cM_{ab} \cM_{ab}, \nn\\
C^2[\mso(5)] &= \sum_{a<b\leq 5} \cM_{ab} \cM_{ab}.
\label{Casimirs-explcit}
\end{align}
Their eigenvalues are obtained from $C^2 = \langle \L,\L+2\rho\rangle =  n_i \langle\L_i,\L_j\rangle  (n_j+2)_j$
where $\rho= \sum_i \L_i$ is the Weyl vector, and $n_i$ are the Dynkin labels 
$\L = \sum n_i \L_i$. 
This gives
\begin{align}
  C^2[\msu(4)] (n,0,n) &= \frac 14(n,0,n)\begin{pmatrix}
                        3 & 2 & 1  \\
                        2 & 4 & 2 \\
                        1 & 2 & 3
                       \end{pmatrix}\begin{pmatrix}
                                     n+2\\2\\n+2
                                    \end{pmatrix}
 = 2n(n+3) \\
 C^2[\mso(5)] (n-m,2m) &= \frac 12(n-m,2m)\begin{pmatrix}
                         2 & 1\\
                         1 & 1
                        \end{pmatrix}\begin{pmatrix}
                                      n-m+2\\2m+2
                                     \end{pmatrix}
                 = n(n+3) + m(m+1)
 \label{C2-EV}
\end{align}
in agreement with \cite{Medina:2002pc}.
Therefore 
\begin{align}
 \Box (n-m,2m) =  n(n+3) - m(m+1), \qquad m\leq n \ ,
 \label{Box-EV}
\end{align}
which is positive definite as it must be. 
However, note that $\Box$ is rather small for the $(0,2m)$ modes corresponding to $n=m$. 
As a check, we recover $\Box X^i = 4 X^i$ \eq{box-x}.
Finally, we can recover  \eq{radius} from \eq{Casimirs-explcit},
\begin{align}
 \sum_i X_i^2 =  C^2[su(4)] (0,0,N) - C^2[so(5)] (0,N)  = \frac 14 N(N+4) \ .
\end{align}

\section{Fluctuations}

\subsection{Matrix model and mode expansion}

We would like to find the fluctuation modes which arise on a $S^4_N$ background in
the Yang-Mills matrix model action with a mass term,
\begin{align}
 S[X] &= \frac 1{g^2}\Tr \Big(-[X_a,X_b][X^a,X^b]\, + \mu^2 X^a X_a \Big) \ .
 \label{bosonic-action}
\end{align}
The equations of motion are
\begin{align}
 (\Box + \frac 12 \mu^2) X_a  = 0 ,
\end{align}
and \eq{box-x} implies that $S^4_N$ is a solution for $\mu^2=-8$. However
such a negative mass term implies an instability, and it is the purpose of this paper 
to show that quantum effects can overcome this problem. Thus we will only allow 
a positive mass term $\mu^2\geq 0$.

Now we add fluctuations around some (not necessarily on-shell)
background $X^a = \bar X^a + \cA^a(\bar X^a)$.
Expanding the action expanded up to second oder in $\cA^a$, one obtains
\begin{align}
 S[X]  &= S[\bar X]  + \frac{2}{g^2}\Tr \Big(2\cA^a (\Box +\frac 12\mu^2) X_a 
  +\cA_a (\Box +\frac 12\mu^2) \cA_a - 2 [\cA_a,\cA_b] [\bar X^a,\bar X^b] + f^2  \Big) \ . \nn
\end{align}
Here
\begin{align}
  f = [\cA^a,\bar X_a] 
 \label{gauge-fixing-function}
\end{align}
can be viewed as gauge fixing function, which
transforms as 
 $f \to f + \Box \L$
under gauge transformations. 
Hence the quadratic fluctuations $\cA^a$ are governed by the quadratic form
\begin{align}
\Tr \cA_a \Big((\Box + \frac 12 \mu^2)\d^a_b  + 2i [\cM^{ab},. \, ] - [X^a,[X^b,.]]\Big) \cA_b  \ .
\label{fluct-action-nogf}
\end{align}
The last term is canceled upon adding a suitable 
Faddeev-Popov gauge-fixing term for $f=0$ \cite{Blaschke:2011qu}, so that the fluctuations 
are governed by the  ``vector'' (matrix) Laplacian
\begin{align}
(D^2 \cA)_a :=
\big(\Box + \frac 12 \mu^2 - M^{(\cA)}_{rs}[\cM^{rs},.]\big)^a_b   \cA_b \ 
\label{fluct-action}
\end{align}
where 
\begin{align}
(M_{ab}^{(\cA)})^c_d &= i(\d^c_b \d_{ad} - \d^c_a \d_{bd})\, 
\end{align}
is the $SO(5)$  generator  in the vector representation.
A  geometric discussion of such fluctuation modes $\cA$ is given e.g. in
\cite{Steinacker:2012ra}.


\subsection{Mode expansion and diagonalization}

The  fluctuations $\cA_a$ are clearly mixed by the above action, and 
we need the explicit separation into eigenmodes.
From a $SO(5)$ point of view, they live in
\begin{align}
 \cA = v^a\cA_a \in (1,0) \otimes (n,m) 
\end{align}
where $v_a$ denotes a vector  of $SO(5)$.
The intertwiner 
\begin{align}
 \cA_a \to 2i[\cM_{ab},\cA_b] = -\big(M^{(\cA)}_{cd}[\cM^{cd},.]\big)^a_b   \cA_b
\end{align}
has a clear group-theoretical meaning: it is simply $ M_{ab}^{(\cA)} \otimes M_{ab}^{(ad)} $, where
 $M_{ab}^{(ad)} = [\cM_{ab},.]$ is the $\mso(5)$ action on $End(\cH_N)$.
Together with \eq{box-Casimirs}, it follows that $\Box$ commutes with  $M_{ab}^{(\cA)}\otimes M_{ab}^{(ad)}$, 
and we can simultaneously diagonalize those operators. 
Using \eq{Casimirs-explcit},  we simply need to diagonalize 
\begin{align}
M_{bc}^{(ad)} \otimes M_{bc}^{(\cA)}
  &= C^2[\mso(5)]^{(\cA)\otimes(ad)} - C^2[\mso(5)]^{(ad)} - C^2[\mso(5)]^{(\cA)} .
  \label{intertwiner}
\end{align}
The required tensor product decomposition is easily obtained
\begin{align}
 \cA \in (1,0) \otimes (n-m,2m) 
  &= (n-m+1,2m) \oplus  (n-m-1,2m+2)   \oplus  (n-m,2m)   \nn\\
   &\quad  \oplus  (n-m+1,2m-2) \oplus  (n-m-1,2m)  
  \label{mode-decomp-2}
\end{align}
for generic $(n,m)$, which gets truncated for  $m=0$ as follows 
\begin{align}
(1,0)\otimes (n,0)  &= (n+1,0) \oplus (n-1,2) \oplus (n-1,0)  , \qquad n \geq 1 \nn\\
(1,0)\otimes (0,0)  &=(1,0)
\end{align}
and for $n=m \geq 1$
\begin{align}
(1,0)\otimes (0,2m)  &= (1,2m) \oplus(0,2m)    \oplus  (1,2m-2),   \qquad m \geq 1 \ .
\end{align}
Using the eigenvalues of the Casimir \eq{C2-EV},
we can  read off the eigenvalues of the intertwiner \eq{intertwiner}
for $A_a \in  (n-m,2m)$:
\begin{align}
 (M_{bc}^{(ad)} \otimes M_{bc}^{(vect)})|_{\cA\in(n-m+1,2m)} 
   &=2n  \nn\\
 (M_{bc}^{(ad)} \otimes M_{bc}^{(vect)})|_{\cA\in(n-m-1,2m+2)} 
   &=2m -2 \nn\\
(M_{bc}^{(ad)} \otimes M_{bc}^{(vect)})|_{\cA\in(n-m,2m)} &=  - 4 \nn\\
(M_{bc}^{(ad)} \otimes M_{bc}^{(vect)})|_{\cA\in(n-m+1,2m-2)} 
   &= -2m-4 \nn\\
(M_{bc}^{(ad)} \otimes M_{bc}^{(vect)})|_{\cA\in(n-m-1,2m)} 
   &= -2n-6 \ .
  \label{vector-term-eom}
\end{align}
Therefore the eigenvalues of the vector Laplacian $D^2$ 
 for $A_a \in (n-m,2m)$ are given by
\begin{align}
  D^2|_{\cA\in (n-m+1,2m)}  
   &= n(n+1)-m(m+1) + \frac 12 \mu^2   \nn\\
 D^2|_{\cA\in(n-m-1,2m+2)} 
      &= n(n+3)-m(m+3)+2 + \frac 12 \mu^2  \nn\\
 D^2|_{\cA\in(n-m,2m)} &= n(n+3)-m(m+1) + 4 + \frac 12 \mu^2 \nn\\
 D^2|_{\cA\in(n-m+1,2m-2)}
     &= n(n+3)-m(m-1) + 4 + \frac 12 \mu^2  \nn\\
 D^2|_{\cA\in(n-m-1,2m)} 
      &= n(n+5)-m(m+1)+6 + \frac 12 \mu^2 \ .
  \label{vector-Laplace-modes}
\end{align}
For $\mu^2 = 0$, all  modes are positive except for 
the $(n-m+1,2m)$ modes with $n=m$, which are zero modes.
For $n=m=0$ these are the translation modes $\cA_a = c_a$. 
More generally, they are  massless higher spin fields at zero momentum.
For example, the  $(1,2) \subset (1,0)\otimes (0,2)$ modes are given by $\cA_a = T_{abc}\cM_{bc}$
with $T$ anti-symmetric in the last indices and traceless.
These correspond to excitations of the internal $S^2$.

\paragraph{$SO(5)$ Goldstone bosons.}

As a check, we identify the $SO(5)$ Goldstone modes
$\d X_a = \cA_a = \L_{ab} X_b$ for antisymmetric traceless $\L_{ab}$,
which should be zero modes of \eq{fluct-action-nogf} (before gauge fixing!) for $\mu^2 = -8$. 
These are the $(0,2) \subset (1,0)\otimes (1,0)$  modes
corresponding to 4 translational  and 6 rotational modes.
By construction, these are equivalent to pure gauge modes generated by 
$[\frac 12\L_{ab}\cM_{ab},.]$.
We compute 
\begin{align}
 2i[\cM_{ab},\L_{bc} X_c] &= - 2 \L_{bc} (\d_{ac} X_b - \d_{bc} X_a  ) 
 =  2 \L_{ab}X_b \ .
\end{align}
This cancels with the term 
\begin{align}
 [X^a,[X^b,\L_{bc} X_c]] &= i \L_{bc}[X^a,\cM_{bc}]
   = 2 \L_{ac}X_c \ ,
\end{align}
therefore for $\mu^2 = -8$ these are indeed zero modes of \eq{fluct-action-nogf},
\begin{align}
\Big((\Box + \frac 12 \mu^2)\d^a_b  + 2i [\cM^{ab},. \, ] - [X^a,[X^b,.]]\Big) \cA_b = 0 .
\end{align}

\subsection{Fermions}
\label{sec:fermions}

Fermions are organized similarly. They are governed by the action 
\begin{align}
 S[\Psi] = \Tr \obar\Psi \Gamma_a[X^a,\Psi] \ .
\end{align}
In the Lorentzian IKKT model, the spinors are (matrix-valued) Majorana-Weyl spinors $\Psi$ of $SO(1,9)$, 
in particular
\be
\Psi_C = \cC \obar\Psi^T = \Psi
\,, 
\ee
where\footnote{The transpose in $\Psi^T$ refers only to the 10 spinor indices.} 
$\gamma_a^T = \cC \gamma_a \cC^{-1}$. Accordingly,  
the Grassmann integral over the MW spinors yields 
\begin{align}
e^{i\Gamma^\psi[X]} &=\int d\Psi e^{i\Tr \obar\Psi \Gamma_a[X^a,\Psi]} 
  = {\rm Pfaff}(\tilde\Gamma_a^{\a\b} [X^a,.])= \pm\sqrt{\det(\cC\slashed{D}_+)}
\,, \label{eq:Grassmann-integral-MW-spinors}
\end{align}
where
$\cC\slashed{D}_+$ denotes $\cC\slashed{D}$ acting on the positive chirality spinors. 
This expression makes sense also in the Euclidean case, where
the effective action has in general both real and imaginary contributions. 
The real part of the action can be extracted from
\begin{align}
\det \((\cC \slashed{D})^\dagger \cC \slashed{D}_+\) = \det(\slashed{D}^2_+)
 = e^{-2\Re(\Gamma_E^\psi[X]) } \, .
\end{align}
The imaginary part is the Wess-Zumino contribution, which however vanishes 
on backgrounds with reduced dimensions \cite{Ishibashi:1996xs} such as ours.
Therefore we only need the spectrum of 
\begin{align}
 \slashed{D}^2 = \Box - M^{(\psi)}_{ab}[\cM_{ab},.]
 \label{Dirac-square}
\end{align}
on the 16-dimensional spinor representation of $SO(10)$, where
 \begin{align}
 M_{ab}^{(\psi)} &= \frac{1}{4i}[\G_a,\G_b] \ . 
\end{align}
The spinors live in 
\begin{align}
 \Psi &= (0,1) \otimes (n-m,2m) \nn\\
  &= (n-m,2m+1) \oplus (n-m+1,2m-1)  \oplus (n-m-1,2m+1) \oplus (n-m,2m-1) \nn
\end{align}
where $(0,1)$ is the $SO(5)$ spinor representation.
Note that the second Dynkin index is now odd, the modes being fermions.
This holds for generic $(n, m)$, but gets truncated for  $m=0$ as follows 
\begin{align}
 (0,1) \otimes (n,0) &= (n,1)   \oplus (n-1,1) , \qquad n \geq 1 \nn\\
 (0,1) \otimes (0,0) &= (0,1) 
\end{align}
and for $n=m\geq 1$ as follows 
\begin{align}
 (0,1) \otimes (0,2m) &= (0,2m+1) \oplus (1,2m-1)   \oplus (0,2m-1) \ .
\end{align}
Again we simply need to diagonalize 
\begin{align}
  M^{(\psi)}_{ab}[\cM_{ab},\Psi] &= (M_{ab}^{(ad)} \otimes M_{ab}^{(\psi)}) \Psi \nn\\
  &= \Big(C^2[so(5)]^{(ad)\otimes(spin)} - C^2[so(5)]^{(ad)}-C^2[so(5)]^{(spin)}\Big) \Psi \ .
\end{align}
Using the eigenvalues of the Casimir \eq{C2-EV},
we can  read off the eigenvalues of this intertwiner
for $\Psi \in (0,1)\otimes (n-m,2m)$:
\begin{align}
 (M_{bc}^{(ad)} \otimes M_{bc}^{(\psi)})|_{\Psi\in (n-m,2m+1)} 
   &= n +m \nn\\
 (M_{bc}^{(ad)} \otimes M_{bc}^{(\psi)})|_{\Psi\in (n-m+1,2m-1)} 
   &= n-m-1 \nn\\
(M_{bc}^{(ad)} \otimes M_{bc}^{(\psi)})|_{\Psi\in (n-m-1,2m+1)} 
   &= -n +m-3 \nn\\
(M_{bc}^{(ad)} \otimes M_{bc}^{(\psi)})|_{\Psi\in (n-m,2m-1)} 
   &= -n-m-4  \ .
\end{align}
Therefore the eigenvalues of  $\slashed{D}^2$ \eq{Dirac-square}
 for the components of $\Psi$ in $(n-m,2m)$ are
\begin{align}
 \slashed{D}^2|_{\Psi\in (n-m,2m+1)} 
   &= n(n+2)-m(m+2)  \nn\\
 \slashed{D}^2|_{\Psi\in (n-m+1,2m-1)} 
      &= n(n+2)-(m-1)(m+1) \nn\\
  \slashed{D}^2|_{\Psi\in (n-m-1,2m+1)} 
   &= n(n+4)-(m-1)(m+3)  \nn\\
 \slashed{D}^2|_{\Psi\in (n-m,2m-1)} 
     &= n(n+4)-(m-2)(m+2) \ .
  \label{Diracsquare-modes}
\end{align}
There are again zero modes $(n-m,2m+1)$  for $n=m$, however
with different multiplicity $\dim (0,2m+1)$ compared with \eq{vector-Laplace-modes}. Hence 
supersymmetry is clearly broken, but some traces do remain.

\section{The 1-loop effective potential}
\label{sec:one-loop}

Now we are in a position to compute the 1-loop effective action on the $S^4_N$ background
\begin{align}
 X_i = r \bar X_i, \qquad \sum_i \bar X_i \bar X_i = \frac 14 N(N+4) \one
\end{align}
with a small positive mass term as in \eq{bosonic-action}. This mass $\mu^2$ serves as a ``source'' 
for the radial parameter $r$, which will be determined by the 1-loop effective action 
\begin{align}
 \Gamma_{\!\textrm{eff}}[r,\mu] = S[r,\mu] + \Gamma_{\!\textrm{1loop}}[r,\mu] \ 
\end{align}
defined by
\begin{align}
 Z[r,\mu] = \int\limits_{\rm 1\, loop} dX d\Psi e^{-S[r \bar X,\Psi]} 
 = e^{- \Gamma_{\!\textrm{eff}}[r,\mu] } \ .
\end{align}
We are going to show that 
$\Gamma_{\!\textrm{eff}}[r,\mu]$ has a non-trivial minimum $r=r_0(\mu)$ provided $\mu \neq 0$.
$\mu^2$ also serves to regularize the bosonic zero modes (but not the fermionic ones).
The bare bosonic action \eq{bosonic-action} for the background $X = r \bar X$ is computed as 
\begin{align}
  S[X] &= \frac 1{g^2}\Tr \Big(-[X_i,X_j][X^i,X^j]\, + \mu^2 X^i X_i \Big)  \nn\\
      &= \frac{1}{g^2}\Tr \big( r^4\bar\cM_{ij} \bar\cM_{ij}\, + \frac{\mu^2}4 r^2 N(N+4) \one \big) \nn\\
   &= \frac{1}{6g^2} r^4(1 + \frac 14\tilde \mu^2  )N(N+4)(N + 1)(N + 2)(N + 3) 
\end{align}
using $C^2[\mso(5)] (0,N) = \frac 12 N(N+4)$ and \eq{dim-HN} in the last step, and setting 
\begin{align}
 \tilde \mu^2 = \frac{\mu^2}{r^2} \ .
\end{align}
We recall  
the following form of the one-loop effective action in the 
IKKT model \cite{Ishibashi:1996xs,Chepelev:1997av,Blaschke:2011qu} 
\begin{align}
\Gamma_{\!\textrm{1loop}}[X]\! &= \frac 12 \Tr \Big(\log(r^2\Box +\frac{\mu^2}2 - M^{(\cA)}_{ab}[r^2\bar\cM^{ab},.])
-\frac 12 \log(r^2\Box - M^{(\psi)}_{ab}[r^2\bar\cM^{ab},.])
- 2 \log (r^2\Box)\Big)   \nn\\
 &= \frac 12 \Tr \Bigg(\sum_{n>0} \frac{1}n \Big((\Box^{-1}\big(M^{(\cA)}_{ab}[\bar\cM^{ab},.] 
    - \frac{1}{2}\tilde \mu^2)\big)^n 
  \, -\frac 12 (\Box^{-1}M^{(\psi)}_{ab}[\bar\cM^{ab},.])^n \Big)  \Bigg) 
\label{Gamma-IKKT}
\end{align}
with  $a,b=1,...,10$,
where
\begin{align}
\begin{array}{rl}(M_{ab}^{(\psi)})^\a_\b &= \frac 1{4i} [\gamma_a,\gamma_b]^\a_\b \,  \\
                      (M_{ab}^{(\cA)})^c_d &= i(\d^c_b \d_{ad} - \d^c_a \d_{bd}) \, , \\
                    \end{array}  
\end{align}
and the  $2 \log \Box$ term arises from the ghost contribution.
Here $\Box$ and $\bar\cM^{ab}$ refer to the operators defined for the background $\bar X_i$ 
as in the previous sections.
Note that the coupling constant $g$ drops out  from $\Gamma_{\!\textrm{1loop}}$
due to supersymmetry, and the radius $r$ enters only through $\tilde\mu^2$.
For $\mu=0$, the first non-vanishing term in this expansion is  $n=4$ due to maximal supersymmetry.
However there are contributions of order $\cM$ for $\mu^2\neq 0$ due to the soft SUSY breaking, 
which will be crucial below.

This  1-loop effective action can be written neatly 
in exponentiated form \cite{Blaschke:2011qu} using a Schwinger parametrization
\begin{align}
\Gamma_{\!\textrm{1loop}}[X]\! &= - \frac 12 \Tr \int\limits_0^\infty \frac {ds}{s} 
   \Big( e^{-sr^2(\Box +\frac{1}{2}\tilde \mu^2  - M^{(\cA)}_{ij}[\bar\cM^{ij},.])}
  - \frac 12 e^{-s r^2(\Box -4M^{(\psi)}_{ij}[\bar\cM^{ij},.]) }  - 2 e^{-s r^2\Box}  \Big) \nn\\
&= - \frac 12 \Tr \int\limits_0^\infty \frac {d s}{s}  e^{-s\Box}
   \Big( e^{s (-\frac{1}{2}\tilde \mu^2  +M^{(\cA)}_{ij}[\bar\cM^{ij},.])}
  - \frac 12 e^{sM^{(\psi)}_{ij}[\bar\cM^{ij},.] }  - 2  \Big)
\label{finite-N-IKKT}
\end{align}
noting that $[\Box,[\bar\cM_{ij},.]] = 0$. 
This form will allow us to capture the full 1-loop effective action, and to
justify the truncation at $n=4$  in the expansion  \eq{Gamma-IKKT}.

\paragraph{Flat branes.}
Before evaluating the trace explicitly, let us pause and
try to understand what to expect.
For $\mu^2=0$, the 1-loop effective action is expected to reproduce  
(at least part of) IIB supergravity, in particular the leading supergravity interaction between 
flat D-branes in $\R^{10}$ is known to be recovered \cite{Ishibashi:1996xs,Chepelev:1997av}. 
We recall the case of (not necessarily parallel) flat branes with 2-form fluxes, 
as realized by quantum planes.
For $N$   $\R^4_{\theta_a}$ branes with 
noncommutative structures\footnote{$\theta^{-1}$ corresponds to the 2-form field $B+F$ on the brane in string theory.}
$\theta^{ij}_a, \, a=1,...,N$, 
the trace in \eq{finite-N-IKKT} is over $\cA_\theta \otimes End(V) \otimes \mmu(N)$ where $\cA_\theta$ is the 
algebra of functions on $\R^{4}_{\theta}$ and $V$ stands for the $SO(10)$
vector, spinor or scalar representation. 
We can interpret $\mmu(N)$ as ``strings'' $|a\rangle\langle b|$ connecting brane $a$ with brane $b$, and 
replace $[\bar\cM^{ij},.]$ by $\theta^{ij}_{ab} :=\theta^{ij}_a-\theta^{ij}_b$.
One can then evaluate the trace over the internal $SO(10)$ part explicitly,
which gives \cite{Blaschke:2011qu}
\begin{align}
 tr_{10}(e^{s M^{(\cA)}_{ij}\theta^{ij}_{ab}}) 
 - \frac 14 tr_{32}(e^{sM^{(\psi)}_{ij}\theta^{ij}_{ab} }) - 2
&= (e^{s (f_1-f_2)/2} - e^{-s (f_1-f_2)/2})^2(e^{s (f_1+f_2)/2} - e^{-s (f_1+f_2)/2})^2 \nn\\
 &\geq 0
\label{chi-rank4}
\end{align}
where $\pm f_1, \pm  f_2$ are the eigenvalues of the rank 4 tensors $i\theta^{\mu\nu}_{ab}$, for fixed $a\neq b$. This must  be summed over 
all $N^2-N$ pairs $(a,b)$, corresponding to the trace over $\mmu(N)$.
The crucial point is that this is {\em positive,} leading to an {\em attractive} interaction, which vanishes
only in the (anti-) selfdual case $f_1 = \pm f_2$. 
In particular, parallel $4$-dimensional branes with identical Poisson structure 
are  non-interacting, due to their BPS property.
For $f_1 \neq \pm f_2$, the trace over $\cA$ correctly reproduces the attractive 
$-\frac 1{r^4}$  potential
between flat branes at distance $r$  \cite{Ishibashi:1996xs,Chepelev:1997av}.
On the other hand, the interaction is in general {\em not} attractive for fluxes with rank $\geq 6$, therefore 
the  mechanism under consideration does {\em not} apply to  higher-dimensional spaces. 

\paragraph{Fuzzy $S^4_N$.}

This insight can be applied to the fuzzy 4-sphere, by viewing $S^4_N$  as a stack of $N+1\, $ 
non-commutative 4-spheres with different $\theta^{\mu\nu}_a$ on each sheet.
The short strings connecting these sheets at the same point $p\in S^4$ will not contribute, 
since $\theta^{\mu\nu}_a-\theta^{\mu\nu}_b$ 
 is selfdual as discussed in section \ref{sec:brane-stack}. This is  important, otherwise there would be an 
 instability\footnote{In the flat case, coinciding branes are stable only for (anti-) selfdual gauge fields.}. 
However the longer strings connecting different points with different tangent spaces
will induce a non-vanishing, and therefore attractive, interaction. This means that the sheets in $S^4_N$
are bound to each other, but this alone would not prevent the radius from shrinking since $r$ drops 
out in \eq{finite-N-IKKT} for $\mu^2=0$. 
However for $\mu^2 >0$, the bosonic fluctuations (i.e. strings) are suppressed, while the fermionic ones 
are unaffected. Since the bosonic strings are responsible for the attractive interaction, 
the binding energy is reduced, leading to weaker binding (or even repulsion) 
for larger $\tilde\mu^2= \frac{\mu^2}{r^2}$  due to SUSY breaking.
This means that if $r$ is small, the binding energy is reduced, while for large $r$ that SUSY breaking effect 
becomes irrelevant. 
This is the basic stabilization mechanism at work here, which we will verify in detail below.
Since the above argument can be understood in terms of the relation of the IIB matrix model to IIB supergravity
(which is expected to hold at the  quantum level beyond one loop), 
we expect this mechanism to apply beyond the one loop computation given below.

Now we evaluate the trace in \eq{finite-N-IKKT} using the above mode decomposition. This gives
\begin{align}
 &\Gamma_{\!\textrm{1loop}}[X]\! =- \frac 12 \int\limits_0^\infty \frac {d s}{s} 
 \sum_{n=1}^N\Big(\sum_{m= 1}^{n-1} e^{-s\big(n(n+3)-m(m+1)\big)} \chi_{nm} 
  + e^{-s(n(n+3))} \chi_{n0}  + e^{-s(2n)} \chi_{nn}  \Big)
  \label{1-loop-schwinger-so5}
\end{align}
where the ``generic'' contribution is 
\begin{align}
  \chi_{nm} &= e^{-\frac 12 s\tilde \mu^2}\! \Big(e^{2s n}  \dim(n-m+1,2m) 
  + e^{2s(m-1)} \dim(n-m-1,2m+2) 
   + e^{-4s} \dim(n-m,2m) \nn\\
  & + e^{-2s(m+2)} \dim(n-m+1,2m-2) 
   + e^{-2s(n+3)} \dim(n-m-1,2m)  
  + 5 \dim(n-m,2m) \Big) \nn\\
 & - 2 \dim(n-m,2m)  -2 \Big(e^{s(n+m)} \dim(n-m,2m+1)
   + e^{s(n-m-1)} \dim(n-m+1,2m-1) \nn\\
 &\qquad  + e^{-s(n-m+3)} \dim(n-m-1,2m+1)
   + e^{-s(n+m+4)} \dim(n-m,2m-1) \Big) \nn\\[1ex]
   &= \chi_{nm}^{(0)}(s) + \tilde \mu^2 \chi_{nm}^{(1)}(s) + O(\tilde\mu^4) 
   \label{chinm-expand}
 \end{align}
 where
 \begin{align}  
  \chi_{nm}^{(0)}(s)  &= \dim(n-m,2m) \Big(m (1 + m) (-7 + m + m^2) + 3 n - 6 m (1 + m) n - \nn\\
 &\quad   2 (-5 + m + m^2) n^2 + 6 n^3 + n^4 \Big)s^4 + O(s^5) \nn\\
 \chi_{nm}^{(1)}(s)  &= -5 \dim(n-m,2m) s + O(s^2)  \ .
 \label{chinm-expand-trunc}
\end{align}
The contribution from the $\cA\in (n,0)$ modes is 
\begin{align}
 \chi_{n0} &= e^{-\frac 12 s\tilde \mu^2}\Big(e^{2sn}  \dim(n+1,0) 
  + e^{-2s}  \dim(n-1,2) 
  + e^{-2s(n+3)} \dim(n-1,0) 
  + 5 \dim(n,0)\Big)  \nn\\
  & - 2\dim(n,0) - 2 \Big(  e^{s n} \dim(n,1) + e^{-s(n+3)} \dim(n-1,1) \Big)\nn\\[1ex]
    &= \chi_{nm}^{(0)}(s) + \tilde \mu^2 \chi_{nm}^{(1)}(s) + O(\tilde\mu^4) 
\end{align}
 where
 \begin{align}
  \chi_{n0}^{(0)}(s)  &= \dim(n,0)  n  (n+3)  (1 + n (n+3))\, s^4 + O(s^5) \nn\\
  \chi_{n0}^{(1)}(s)  &= -5 \dim(n,0) s  + O(s^2) \ .
 \end{align}
Finally the contribution from the $\cA\in (m,m)$ modes is 
\begin{align}
 \chi_{mm} &= e^{-\frac 12 s\tilde \mu^2}\Big( e^{2sm} \dim(1,2m)
   + e^{-4s} \dim(0,2m)
   + e^{-2s(m+2)}\dim(1,2m-2) \nn\\
    & \quad + 5 \dim(0,2m) \Big) -2\dim(0,2m) \nn\\
   &  - 2 \Big( e^{2s m} \dim(0,2m+1)
   + e^{-s} \dim(1,2m-1) + e^{-2s(m+2)} \dim(0,2m-1)  \Big) \nn\\[1ex]
     &= \chi_{mm}^{(0)}(s) + \tilde \mu^2 \chi_{mm}^{(1)}(s) + O(\tilde\mu^4) \
\end{align}
where
\begin{align}
  \chi_{mm}^{(0)}(s)  &= -2 \dim(0,2m) m (m+2) s^4 +  O(s^5)  \nn\\
  \chi_{mm}^{(1)}(s)  &= -5\dim(0,2m)  s  + O(s^2) \ .
\end{align}
As a check, we note that  all terms start at $O(s^4)$ for $\mu=0$. 
This corresponds to the fact that the expansion  \eq{Gamma-IKKT} starts at $n=4$ due to maximal supersymmetry.
Furthermore, we observe that all the generic $\chi_{nm}$ and  $\chi_{n0}$ 
contributions  to $\Gamma_{\!\textrm{1loop}}$ are  negative.
This reflects the attractive interaction between the branes as discussed above, cf. \eq{chi-rank4}.
In contrast, the $\mu^2\neq 0$ contributions to  $\Gamma_{\!\textrm{1loop}}$ are positive,
as  the fermionic contributions start to dominate.

The $\chi_{mm}$ appear to give a positive (repulsive) 
contribution; however these  contain the zero modes, which should  not be integrated out,
and we will drop them from now on. 
One may hope that most of  them acquire a mass due to quantum effects; this  
 should be studied in  detail elsewhere. 
In any case,  the ''generic`` modes with $1\leq m \leq n-1$ will give the dominant contribution  for large $N$.
This leads to the desired effective action which governs  
the radius $r$ of $S^4_N$, as elaborated below.

\begin{figure}
\begin{center}
 \includegraphics[width=0.55\textwidth]{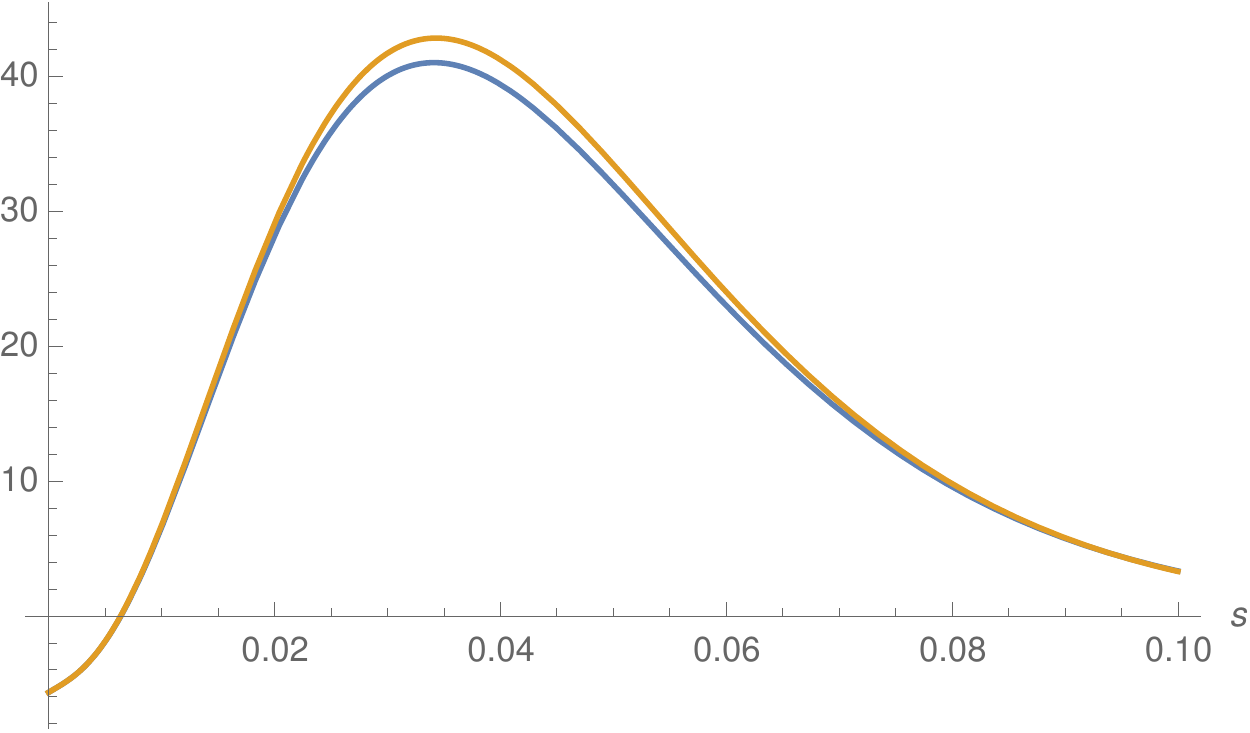}
 \end{center}
 \caption{$\frac 1s\, e^{-s(n(n+3)-m(m+1))}\chi_{nm}(s)$ versus  
 $\frac 1s\, e^{-s(n(n+3)-m(m+1))}(O(s^4) + \tilde\mu^2 O(s))$ as in \eq{chinm-expand-trunc}
 for $n = 9, m = 4, \mu = 0.02$. The negative contribution for small $s$ arises from SUSY breaking 
 due to  $\mu\neq 0$.}
 \label{fig:chinm-expand}
\end{figure}

It is now easy to see that replacing $\chi_{nm}$ by their leading 
$O(s^4) + \tilde\mu^2 O(s)$ expansion as in the last line of \eq{chinm-expand} is a very good approximation, 
as long as $n>m$ and $N$ is large. 
The reason is that the small $s$ regime provides the dominant contribution, due to the
suppression from $e^{- s \Box}$; this is illustrated numerically in figure \ref{fig:chinm-expand}.
We will therefore keep only this leading expansion\footnote{Thus from now on we are basically 
back to  \eq{Gamma-IKKT}, but the above  discussion provides the justification for 
this truncation, as well as a neat way to evaluate it.}, which gives the leading contribution for large $N$.
Then the integral over $s$ can be trivially evaluated.
To perform the sum\footnote{It is easy to see that the contribution from $\chi_{n0}$
is correctly recovered by including $m=0$ in that sum. 
However, this does not make any difference for large $N$.} $\sum\limits_{n=1}^N\sum\limits_{m= 1}^{n-1}$,
we can make the following simplifications which are valid for large $n,m$:
\begin{align}
  \chi_{nm}(s)  \ &\sim  \ 
   \dim(n-m,2m) \big( (n^2 - m^2 )^2 s^4 -5 \tilde\mu^2 s \big) \nn\\
 \dim(n-m,2m)\ &\sim \ \frac 23 n m  (n^2-m^2) 
\end{align}
and 
\begin{align}
 \frac 1{(n(n+3)-m(m+1))^4} \ \sim \ \frac 1{(n^2-m^2)^4} \ .
\end{align}
Approximating the sum by an integral for large $N$, we obtain
\begin{align}
  \Gamma_{\!\textrm{1loop}} \ 
   &\stackrel{N\to\infty}{\sim} \  -\frac 12 \int_0^N dn \int_0^{n-1} dm 
   \Big(\frac{4n m }{n^2-m^2}  - \frac{10}{3}n m\tilde \mu^2 \Big)\nn\\
   &\  \sim \ \ - \frac 12 N^2 \ln N + \frac 5{24} N^4 \tilde \mu^2 \ .
 \label{Gamma-loop-leading}
\end{align}
Here the singularity  at $n=m$ (due to the modes with low masses)
is just barely avoided,
and the main contribution comes from the far UV region  $m \lesssim n$.
We can check the validity of this approximation by
keeping the exact form of  the $O(s^4)+\tilde \mu^2 O(s)$ terms 
in $\chi_{nm}$ \eq{chinm-expand-trunc}.
This gives 
\begin{align}
  \Gamma_{\!\textrm{1loop}}\ &\sim \ 
 - \frac 12 N^2 \left(\ln N-\frac 54- \ln 2\right) + O(N \ln N)\
  + \big(\frac 5{24} N^4 + O(N^3) \big) \tilde \mu^2
  \nn\\
   &\stackrel{N\to\infty}{\sim}  -\frac 12 N^2  \ln N + \frac{5 }{24} N^4\,  \frac{\mu^2}{r^2}
\end{align}
in agreement with \eq{Gamma-loop-leading}.
Thus for $\mu=0$, the one-loop contribution gives indeed an attractive  potential 
$\Gamma_{\!\textrm{1loop}}\sim -\frac 12 N^2  \ln N$, and the factor $N^2$ clearly 
reflects the $N(+1)$ coinciding branes which constitute $S^4_N$.
However  for $\mu^2>0$, the last term describes
a strong repulsive fermionic contribution $\Gamma_{\!\textrm{1loop}} > 0$ 
due to  SUSY breaking\footnote{The higher $O(\mu^4)$ terms are easily seen to be negligible.}.
In particular, we see that the attractive contribution 
dominates for $r\to \infty$, while the repulsive contribution dominates for small $r$
as anticipated.

\begin{figure}
\begin{center}
 \includegraphics[width=0.55\textwidth]{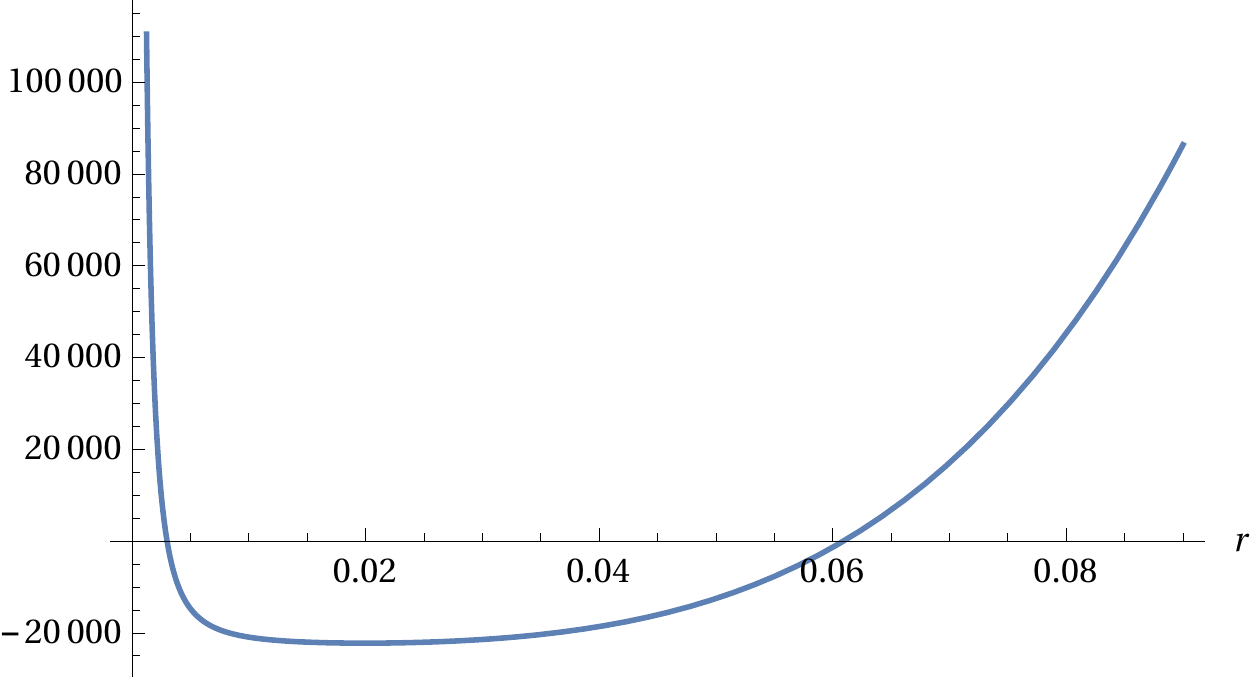}
 \end{center}
 \caption{One-loop effective potential $\Gamma_{\rm eff}[r]$ for $\mu=10^{-4}, g=1, N=100$.}
 \label{fig:potential}
\end{figure}

To finally demonstrate the stabilization of $r$, consider the full one-loop effective potential 
\begin{align}
\Gamma_{\rm eff}[r]  
 \ \sim \ \frac{1}{6g^2} r^4(1 + \frac{1}{4}\tilde \mu^2  )N^5
  -\frac 12 N^2  \ln N + \frac 5{24} N^4 \tilde \mu^2
 \label{potential-S4}
\end{align}
for large $N$, cf. figure \ref{fig:potential}.
For given $\mu^2>0$, 
this has a unique stable minimum  $r=r_0$  at 
\begin{align}
  \frac{r^6}{\mu^6}(1 + \frac{\mu^2}{8r^2})  &= \frac{5}{8}  \frac{g^2}{\mu^4} \frac 1N
\end{align}
while for $\mu^2=0$ the minimum is aways at $r=0$.
The parameter 
\begin{align}
 \epsilon := \frac{5}{8N}  \frac{g^2}{\mu^4} 
 \label{epsilon-def}
\end{align}
determines two scaling regimes:
\begin{align}
\und{\epsilon \gg 1}: &  \qquad  \frac{r^6}{\mu^6} \sim \epsilon \gg 1, \qquad 
     r \sim  \big(\frac{5g^2 \mu^2}{8N}\big)^{\frac 16} \ \  
\qquad  \qquad \mbox{Yang-Mills dominated} \nn\\
\und{\epsilon \ll 1}: & \qquad \frac{r^4}{\mu^4} \sim 8\epsilon \ll 1, \qquad 
     r \sim  \big(\frac{5g^2}{N}\big)^{\frac 14} \quad  \qquad \qquad \mbox{Mass dominated} 
 \label{r-eps}
\end{align}
We are mainly interested in the Yang-Mills case $\epsilon \gg 1$ which arises for $\mu \to 0$.

Assuming that $\epsilon$ and $g$ is fixed, we can rewrite \eq{epsilon-def} as
\begin{align}
 \mu &\sim \Big(\frac{5 g^2}{8}\frac 1{N\epsilon}\Big)^{\frac 14} \sim N^{-\frac 14} \ .
\end{align}
Together with \eq{radius}, 
this means that the ``physical'' radius $R = \frac 12 r N$ of $S^4_N$ (i.e. the largest eigenvalues) scales as 
\begin{align}
R:= \|  X^a \| \sim \frac 12\left\{\begin{array}{ll}
                         (\frac 58 \l)^{\frac 14} \epsilon^{-\frac 1{12}}, & \epsilon \gg 1 \\[1ex]
                         (5 \l)^{\frac 14}, & \epsilon \ll 1 
                       \end{array} \right.
 \label{radius-full}
\end{align}
for given $g, \mu^2$, where 
\begin{align}
 \l := g^2 N^3 \sim 6 g^2 \cN
\end{align}
is essentially the t'Hooft coupling of the matrix model, where $\cN = \dim \cH_N$ is the dimension of the matrices \eq{dim-HN}. 
Thus the volume $R^4$ of the four-sphere scales with the t'Hooft coupling $\l$, with a correction factor 
$\epsilon^{-\frac 1{3}}$ if $\epsilon \sim \frac{\l}{(N\mu)^4} \gg 1$. Therefore 
a large sphere typically arises for large $\l$, provided $\mu > 0$.
It is interesting to recall here the AdS/CFT statement that the supergravity regime 
is applicable for large t'Hooft coupling\footnote{Since  $\l$ is the matrix model t'Hooft coupling
while the usual AdS/CFT  statement is about the gauge theory t'Hooft coupling, this argument is 
somewhat illegitimate. See however the discussion following \eq{gYM}.}; 
this also  supports the validity
of our calculations, since the crucial 1-loop effect can be interpreted in terms of supergravity
as pointed out above.

Let us evaluate the effective potential at the minimum $r=r_0$. For $\epsilon \gg 1$, it is 
given by
\begin{align}
 \Gamma_{\rm eff}[r_0] \ &\stackrel{\epsilon\gg 1}{\sim}  \
  \frac{1}{6g^2} r_0^4 N^5  + \frac 5{24} N^4 \frac{\mu^2}{r_0^2} - \frac 12 N^2  \ln N \nn\\
  & \ \sim \frac{5}{16} N^4 \epsilon^{-\frac 13}  - \frac 12 N^2  \ln N 
  \label{potential-S4-min}
\end{align}
using \eq{r-eps} and  \eq{epsilon-def}.
This maybe positive or negative; a negative energy corresponding to a stable minimum arises if
\begin{align}
  \epsilon & > N^{6}(\ln N)^{-3}
\end{align}
(roughly). In view of \eq{radius-full}, this is compatible with the regime of large spheres, 
e.g. for $\mu\sim N^{-2}$ and $g\sim 1$ we have $\epsilon \sim N^7$, 
$R \sim N^{1/6}$ and $\Gamma_{\rm eff}[r_0]<0$.

For  $\epsilon \ll 1$, the effective potential  at the minimum is
\begin{align}
 \Gamma_{\rm eff}[r_0] \ &\stackrel{\epsilon\ll 1}{\sim}  \
   \frac{1}{24g^2} r^2_0\mu^2 N^5  + \frac 5{24} N^4 \frac{\mu^2}{r_0^2} - \frac 12 N^2  \ln N \nn\\
  & \ \sim \frac{5}{24} N^4 (8\epsilon)^{-\frac 12}  - \frac 12 N^2  \ln N 
\end{align}
This is always positive since $\epsilon \ll 1$ by assumption.

An example for a potential with a negative energy minimum is shown in  figure \ref{fig:potential}
for $\mu \ll 1$. 
This means that the extended $S^4_N$ background is preferred over the trivial vacuum. 
Since the mechanism is expected to apply also in the Minkowski model
due to the IR regularization by an imaginary mass term, 
this might explain the spontaneous generation of 3+1-dimensional geometries found in numerical 
simulations of the Minkowskian model \cite{Kim:2011cr}.

\subsection{Discussion}

The crucial point of the stabilization mechanism is that  for small $r$,
quantum corrections in the presence of $\mu^2>0$
induce a strong repulsive (positive) potential due to  SUSY breaking.
For large $r$, SUSY becomes almost exact, and the quantum corrections induce an
attractive (negative) contribution independent of $r$, 
in addition to  the positive  bare potential $\sim O(r^2) + O(r^4)$.
This leads to a stable minimum at $r_0>0$.
For $\mu=0$, there is no stabilization at one loop, and the sphere would collapse. 
This may be (part of) the reason why  4-dimensional geometries 
have not been observed in Monte-Carlo-simulations 
of the Euclidean IIB model without mass term \cite{Nishimura:2011xy}.

Let us try to assess the validity of the present results. First of all,
figure \ref{fig:potential}  makes clear that the basic mechanism is  very robust at one loop. 
There are two important ingredients: 1) $\Gamma_{\rm eff} \to \infty$ as $r\to 0$, and 
2) $\Gamma_{\rm eff}  \to \infty$ as $r\to \infty$.
Both statements are very robust: for $r\to 0$, the mass term $\mu^2 X^2$ becomes dominant and 
leads to a strong deviation from supersymmetry, so that the full power of UV 
divergences kicks in and induces a huge vacuum energy 
(corresponding to the cosmological constant). 
On the other hand for $r\to\infty$, the background approaches a stack of $N$ flat branes, 
with a self-dual background flux. 
Then quantum corrections are expected to be mild, 
so that the bare action dominates and contributes $r^4 N^5$. These two effects should 
be dominant even in the full quantum  theory. 
Therefore the  stabilization mechanism is  robust.

It is important to see that for large radius, the mechanism is clearly local, 
and the bare ``brane tension'' is balanced by the negative vacuum energy;
the breaking of translational invariance by the $X^2$ term in the bare action is then insignificant.
This means that the same mechanism should also apply to other geometries, as long as 
the curvature is not too large. This is reflected in figure \ref{fig:potential},
which shows that
the effective potential is remarkably flat near the minimum, for large $N$.
This means that the precise geometry of the brane is not important.  Thus
geometric deformations are {\em not} suppressed, which is very interesting from the gravity point of view. 
This will be addressed in future work.

\paragraph{Higher order corrections.}

To assess the validity of the one-loop computation, it is useful to 
view the background as a stack of $N$ spherical noncommutative spheres with ``local'' 
Poisson structure $\theta^{\mu\nu}$. Then the fluctuation modes can be viewed 
in terms of $\cN=4$  $U(N)$ noncommutative 
Super-Yang-Mills gauge theory with effective coupling constant \cite{Steinacker:2010rh}
\begin{align}
 \frac 1{4g_{YM}^2} = \frac 1{(2\pi)^2 g^2} {\rm Pf} \theta^{\mu\nu} \sqrt{\det g_{\mu\nu}}
 \ \sim \  \frac 1{(2\pi)^2 g^2} r_N^2 \sim \frac{N^2}{g^2}
 \label{gYM}
\end{align}
using ${\rm Pf} \theta^{\mu\nu}\sim r_N^2$ \eq{Pfaff-estimate} in the $x^\mu$ coordinates; 
note that the scaling factor $r$ drops out here. 
Hence the higher-order perturbative 
corrections boil down to a calculation in $\cN=4$
noncommutative $U(N)$ SYM theory, 
with  t'Hooft coupling $g_{YM}^2N \sim g^2/N$.
Therefore the $k$-loop contributions to the potential are expected to be of order
$N^2 (g^2/N)^{k-1}$.
These generic results are in fact significantly suppressed 
due to maximal supersymmetry, cf. the discussion in \cite{Bal:2004ai} for the case of fuzzy tori.
In any case, for $g=O(1)$ the 2- and higher-loop contributions 
are clearly smaller than the one-loop effective action \eq{potential-S4},
which is at least of order $O(N^2)$.
Therefore we expect that the one-loop results of this paper receive only small 
perturbative corrections for large $N$.
This makes sense, since then the semi-classical geometrical picture applies,
so that the semi-classical evaluation of the (super)gravity interaction should be justified. 
However, due to the (parametrically small) mass parameter 
$\mu^2$ and the associated SUSY breaking,
these arguments are only heuristic, and a more careful consideration of 
possible UV contributions should be given eventually.

Nonperturbatively, the issue is of course much more complicated due to the many possible 
geometric configurations in the matrix model, as illustrated in the next section.
Nevertheless, it seems plausible that for large $N$ where the background becomes 
almost BPS, the ``decay barrier'' of the geometry should indeed be large.

\subsection{One-loop potential for stacks of fuzzy spheres}
\label{sec:fuzzyS2}

The above mechanism is clearly quite general and applies to many similar backgrounds.
For comparison, we repeat the computation for a background of $k$ coinciding fuzzy 
2-spheres $S^2_N$,
given by\footnote{I would like to thank J. Zahn for related collaboration.}
\begin{align}
 X_i = r \bar X_i, \qquad [\bar X_i,\bar X_j] = i \varepsilon_{ijk} \bar X_k,
 \qquad \sum_{i=1}^3 \bar X_i \bar X_i = \frac 14 N(N+2) \ .
\end{align}
The decomposition of the algebra of functions on $S^2_N$ is
 $\cA = \oplus_{n=0}^{N-1} (2n)$, where $(n)$ denotes the highest weight representation 
with Dynkin label $n\in\N$ and spin $2j=n$.
The Casimir is (keeping the same notation as for $S^4_N$)
\begin{align}
 C^2[\mso(3)](n) &= \sum_{a<b\leq 3} \cM_{ab} \cM_{ab} (n) = \frac 14 n(n+2) \ .
\label{Casimirs-explcit-su2}
\end{align}
The bosonic modes have the following (generic) tensor product decomposition 
\begin{align}
 \cA &\in (2) \otimes (n) = (n+2) \oplus (n) \oplus (n-2) 
\end{align}
and for $\cA_i \in  (n)$ we have
\begin{align}
 (M_{bc}^{(ad)} \otimes M_{bc}^{(vect)})|_{\cA\in(n+2)} &=n  \nn\\
 (M_{bc}^{(ad)} \otimes M_{bc}^{(vect)})|_{\cA\in(n)}  &= -2 \nn\\
(M_{bc}^{(ad)} \otimes M_{bc}^{(vect)})|_{\cA\in(n-2)} &= -n - 2 \ .
\end{align}
The fermionic modes have the following (generic) tensor product decomposition 
\begin{align}
 \Psi &\in (1) \otimes (n) = (n+1) \oplus (n-1)
\end{align}
and for $\Psi \in (1)\otimes (n)$ we have
\begin{align}
 (M_{bc}^{(ad)} \otimes M_{bc}^{(spin)})|_{\Psi\in(n+1)} &=\frac n2  \nn\\
(M_{bc}^{(ad)} \otimes M_{bc}^{(spin)})|_{\Psi\in(n-1)} &= -\frac{n+2}{2}\ .
\end{align}
Then the  1-loop action \eq{1-loop-schwinger-so5} is replaced by 
\begin{align}
 &\Gamma_{\!\textrm{1loop}}[X]\! = - \frac 12 \int\limits_0^\infty \frac {d s}{s} 
 \sum_{n=1}^{N-1} e^{-\frac 14sn(n+2)} \chi_{n} 
\end{align}
where the ``generic'' contribution is 
\begin{align}
  \chi_{n} &= \Big(e^{s n}  \dim(n+2) 
  + e^{s(-2)} \dim(n) 
   + e^{-s(n+2)} \dim(n-2) 
  + 7 \dim(n) \Big) e^{-\frac 12 s\tilde \mu^2} \nn\\
 & - 2 \dim(n)  -4 \Big(e^{\frac 14s(2n)} \dim(n+1)
   + e^{\frac 14s(-2n-4)} \dim(n-1)\Big) \nn\\[1ex]
   &= \chi_{n}^{(0)}(s) + \tilde \mu^2 \chi_{n}^{(1)}(s) + O(\mu^4) 
   \label{chinm-expand-S2}
 \end{align}
 with
 \begin{align}  
  \chi_{n}^{(0)}(s)  &= \frac 1{48} n(n+1)(n+2)( 3n^2+6n -4) s^4 + O(s^5) \nn\\
 \chi_{n}^{(1)}(s)  &= - 5 (n+1) s + O(s^2)  .
 \label{chinm-expand-trunc-S2}
\end{align}
Keeping the $O(s^4) +   \mu^2O(s)$ terms as before and 
carrying out the sum, we obtain 
\begin{align}
  \Gamma_{\!\textrm{1loop}}[S^2_N]\ &\sim \
 - \frac {56}{9} + 10  \frac{\mu^2}{r^2}\,\ln N
\end{align}
in the large $N$ limit. This gets multiplied by $k^2$ on a stack of $k$ coincident $S^2_N$.
The bare bosonic action \eq{bosonic-action} for the background $X = r \bar X\otimes \one_k$ is  
\begin{align}
  S[X] 
   = \frac{1}{g^2} r^4(2 + \tilde \mu^2  )N^2(N + 2)k 
\end{align}
and we obtain the full one-loop large $N$ effective potential 
\begin{align}
\Gamma_{\rm eff}[r] &=  \frac{1}{g^2} r^4(2 + \tilde \mu^2  )N^3k  
+  \big(- \frac {56}{9} + 10  \frac{\mu^2}{r^2} \ln N \big) k^2 \ .
\label{S-k-spheres}
\end{align}
For given $\mu^2>0$, 
this has a unique stable minimum  $r=r_0$  at 
\begin{align}
 \frac{r^6}{\mu^6} (1+  \frac{\mu^2}{4r^2}) &= \frac 52 \frac{k\ln N}{N^3} \frac{g^2}{\mu^4}  
\end{align}
while for $\mu^2=0$ the minimum is aways at $r=0$.
Now the parameter 
\begin{align}
 \tilde\epsilon :=  \frac 52 \frac{k\ln N}{N^3} \frac{g^2}{\mu^4}
 \label{epsilon-def-S2}
\end{align}
determines two scaling regimes:
\begin{align}
\und{\tilde\epsilon \gg 1}: &  \qquad  \frac{r^6}{\mu^6} \sim \tilde\epsilon \gg 1, \qquad 
     r \sim  \big(\frac{5g^2 \mu^2}{2}\frac{k\ln N}{N^3}\big)^{\frac 16} \ \  
\qquad  \qquad \mbox{Yang-Mills dominated} \nn\\
\und{\tilde\epsilon \ll 1}: & \qquad \frac{r^4}{\mu^4} \sim 4\tilde\epsilon \ll 1, \qquad 
     r \sim  \big(10 g^2\frac{k\ln N}{N^3}\big)^{\frac 14} \quad  \qquad \qquad \mbox{Mass dominated} 
     \label{r-eps-S2}
\end{align}
We can evaluate the effective potential at the minimum. For $\tilde\epsilon \gg 1$, it is 
given by
\begin{align}
 \Gamma_{1-loop} &\stackrel{\epsilon\gg 1}{\sim}  \frac{2}{g^2} r^4 N^3k 
 +  \big(- \frac {56}{9} + 10 \ln N  \frac{\mu^2}{r^2} \big) k^2 \nn\\
  &\sim 5\epsilon^{-\frac 13} k^2 \big(1 + 2  \ln N \big) - \frac {56}{9}  k^2
\end{align}
using \eq{r-eps-S2} and  \eq{epsilon-def-S2},
which is negative if
\begin{align}
  \tilde\epsilon & > 8(1+2\ln N)^{3}
\end{align}
(roughly). 
For  $\tilde\epsilon \ll 1$, the effective potential  at the minimum is 
always positive for the same reason as in $S^4_N$.

Comparing with \eq{potential-S4-min}, we see that for some parameter range
(e.g. as in figure \ref{fig:potential} and $k=1$), $S^4_N$  has indeed 
a lower one-loop potential than a single fuzzy $S^2$.
However, it is easy to see from \eq{S-k-spheres} that if we vary $k$ for fixed 
dimension $\cN = k N$, then $\Gamma_{\rm eff}$ for $S^2_N$
is minimized for large $k$. 
This means that the quantum number $N$ for a single fuzzy sphere is 
preferred to be small, presumably\footnote{For very small $N$, some of these formulae 
strictly speaking no longer apply; however the main conclusion that large $k$ 
is preferred is unchanged. This argument would presumably also apply to $S^4_N$ to some extent.} 
$N=2$, while $k$ is large. This would suggest that the semi-classical geometries with large $N$ 
may ultimately be unstable. On the other hand, the 1-loop approximation 
is more reliable in the large $N$ case where the background becomes 
almost BPS, and the ``decay barrier'' is expected to be large. These issues are left for 
future investigations.

\section{Conclusion}

We have performed a detailed one-loop computation for fuzzy $S^4_N$
in the IIB matrix model with a (small) positive mass $\mu^2$, and determined 
the one-loop effective potential for the radius $r$.
We have found a robust stabilization 
mechanism for  $S^4_N$ and similar fuzzy spaces, as long as $\mu^2> 0$. 
This can be understood in terms of a negative contribution 
to the potential attributed to supergravity, 
balanced by a positive contribution due to the SUSY breaking effect of $\mu^2$.
The latter is important only for small $r$, leading to a robust stabilization mechanism. 
For suitable parameters the radius becomes large, 
and  the  energy can be negative for small $\mu^2$.
We also argue (somewhat superficially) that higher-order perturbative corrections 
should be small for large $N$. This suggests that large, semi-classical 
spheres should indeed exist as meta-stable configurations in the model.
However, the mechanism is not restricted to the fuzzy $S^4_N$ geometry.

This result is very interesting in the context of recent numerical results on the 
genesis of 3+1-dimensional space-time in the Minkowskian IIB model \cite{Kim:2011cr}.  
Our  mechanism is expected to work also in the Minkowski case, and it 
might help to explain and to interpret these numerical results. 
In fact, an IR regularization of the IIB model is necessary in the Minkowski case, 
which can be interpreted \cite{Kim:2011cr} as a Wick rotation of the present mass term, 
ensuring also the Feynman $i\varepsilon$ prescription. This motivates to introduce  a 
(parametrically small) mass term also in the Euclidean model. It would  
therefore be very interesting 
to reconsider the Euclidean model numerically in the presence of such a mass term.
Even if no spontaneous generation of 4-dimensional spaces might happen due to non-perturbative effects, 
the present background should at least be metastable.

At the non-perturbative level, the situation is clearly much more complicated.
We consider  the case of $k$ coinciding fuzzy 2-spheres at one loop, where the same mechanism 
applies in principle. It turns out that configurations with many small spheres are 
preferred at the one-loop level. However, this result is not 
expected to be reliable beyond one loop.

We also find that the effective potential for the radius of $S^4_N$ is 
remarkably flat near the minimum, for large $N$. 
This means that geometric deformations are {\em not} suppressed, 
which strongly suggests that some type of gravity with massless modes should arise on the 
background. Due to the manifest Lorentz- (or rather Euclidean) invariance, 
the emergence of 4-dimensional general relativity on such backgrounds
in matrix models seems natural, cf.  \cite{Steinacker:2010rh,Heckman:2014xha}.
Since the vacuum energy is fully incorporated (in fact it is {\em the} 
essential ingredient of the stabilization mechanism), one is tempted to speculate that 
this might shed light on the mysteries of dark energy and the cosmological constant. 
However, this can only be addressed in a meaningful way
once the fluctuation modes on $S^4_N$ and their physical significance are understood.
These include in particular a tower of massless higher spin modes, whose 
fate remains to be determined. All these are interesting topics for further work.

\paragraph{Acknowledgements.}

I would like to thank T. Chatzistavrakidis, D. O'Connor, M. Hanada, S. Ramgoolam, 
A. Tsuchiya and C-S. Chu for useful discussions, 
and J. Zahn and J. Karczmarek for related collaboration and correspondence.
This work was primarily supported by the Austrian Science Fund (FWF) grant P24713, and 
in part by the Action MP1405 QSPACE from the European Cooperation in Science and Technology (COST).

\appendix

\section{Semi-classical geometry: $\C P^3$ as $S^2$ bundle over $S^4$}
\label{sec:appendix-Hopf}

The semi-classical geometry of $S^4_N$ is obtained by recognizing  
\eq{JS-S4-full} as  fuzzy version of the  Hopf map
\begin{align}
 X^i \sim x^i: \quad \C P^3 \to S^4 \subset \R^5
 \label{fuzzy-Hopf}
\end{align}
which is defined as follows.
We  view  $\C^4$ as fundamental representation of $SU(4)$.
Acting on a reference point $z^{(0)} =(1,0,0,0)\in \C^4$, $SU(4)$ sweeps out 
the 7-sphere $S^7 \subset \R^8\cong \C^4$.
We can then define the Hopf map 
\begin{align}
 S^7 &\to S^4 \ \subset \R^5 \\
 z^\a &\mapsto x_i = z_\a^* (\g_i)^\a_{\ \b} z^\b \equiv \langle z| \g_i |z\rangle
 \label{Hopf}
\end{align}
where $\g_i$ are the $\mso(5)$ gamma matrices.
It is easy to verify\footnote{e.g. by noting that $\sum_{i}\g_i\otimes\g_i$ acting on $(4)\otimes_S (4)$
is proportional to $\one$, hence we can evaluate $R^2$ e.g. at $z_\a = (1,0,0,0)$.} 
that $R^2 = \sum_{i=1}^5 x_i^2 = 1$, so that the rhs is indeed in $S^4$. 
Since the overall phase of $z_i$ drops out in 
 \eq{Hopf}, this  defines a map 
of $\C P^3  \cong S^7/U(1)$ into $S^4$.
It is thus useful to re-interpret \eq{Hopf} as
\begin{align}
x_i:  \quad \C P^3  \ &\to S^4 \subset \R^5   \nn\\
   |z\rangle\langle z| &\mapsto \langle z| \g_i |z\rangle = tr( |z\rangle\langle z| \g_i) \ .
 \label{Hopf-cp3}
\end{align}
Here  $|z\rangle\in \C^4$, and $\C P^3$ is identified with the space of rank one projectors 
$|z\rangle\langle z|$.
Using $\g_5 = \begin{pmatrix}
              \one_2 & 0\\
              0 & -\one_2
             \end{pmatrix}$ in the Weyl basis,
we have $x_i(p) = (0,0,0,0,1)$ at the reference point $p \in S^4$, with stabilizer
\begin{align}
 H = \{h; [h,\g_5] = 0\} \   \subset SO(5) 
\end{align}
given by 
$$
H \cong SU(2)_R\times SU(2)_L  = SO(4) 
$$
where $SU(2)_L$ acts on the $+1$ eigenspace of $\g_5$. 
The fiber over $p \in S^4$ is determined by
\begin{align}
 \langle z| \gamma^5 |z\rangle = 1 ,
\end{align}
which using the explicit form of
$\g_5$ is given by
\begin{align}
 |z_1|^2 + |z_2|^2 = 1 \ .
\end{align}
This defines $S^3$, which modulo the $U(1)$ phase (passing to $\C P^3$) reduces to $S^2$.
Hence $\C P^3$ is an $S^2$-bundle over $S^4$, and
the $S^2$ fiber over $z^{(0)}$ is obtained by acting with $SU(2)_L$ on $z^{(0)}$.
In contrast, $SU(2)_R \times U(1) \subset H$ is in the stabilizer of $z^{(0)}$, hence it acts trivially on
$\C P^3$.

\begin{figure}
\begin{center}
 \includegraphics[width=0.4\textwidth]{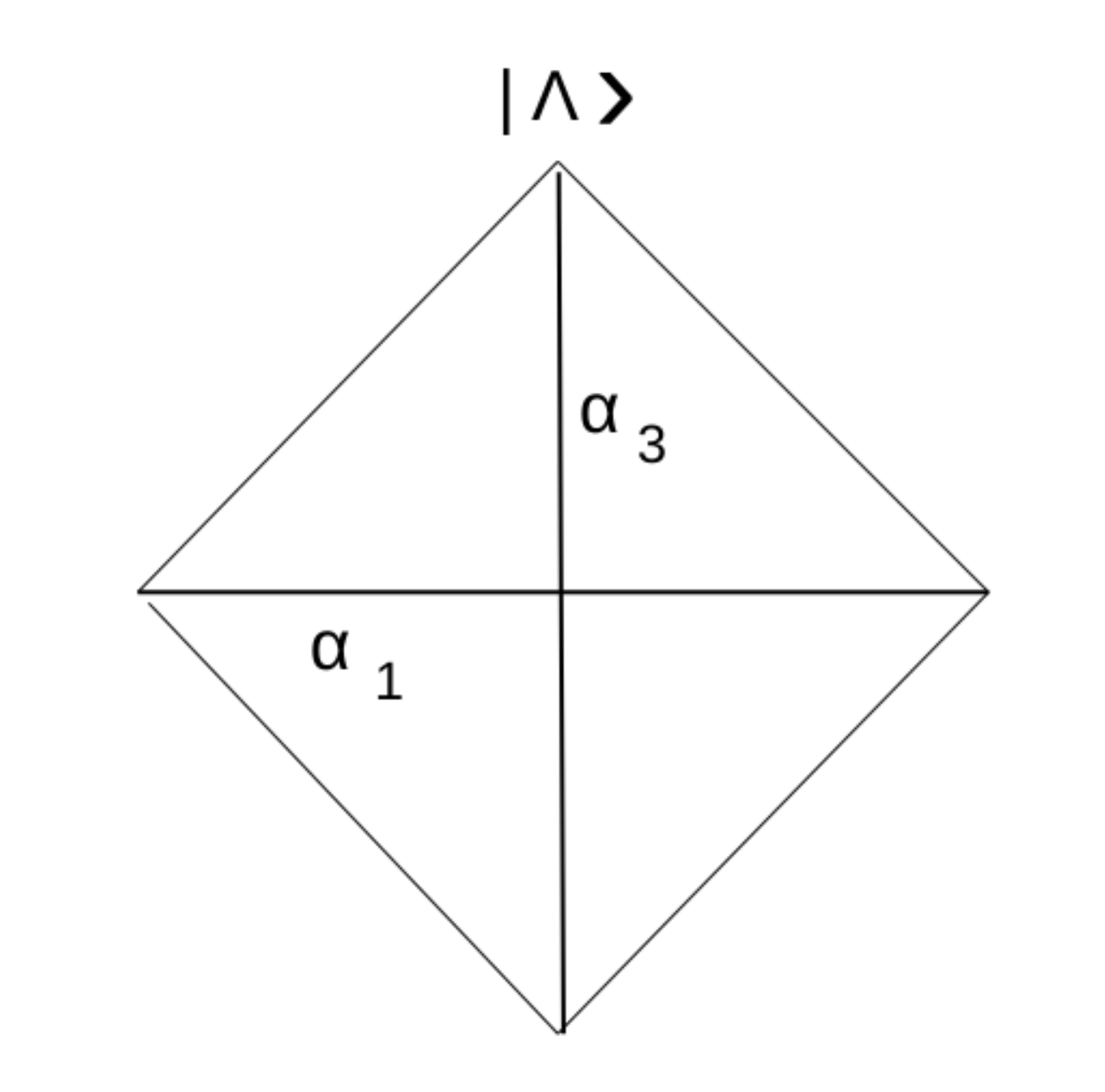}
 \end{center}
 \caption{Weight tetrahedron of $\cH_N$ with highest weight
 $\L = (0,0,N)$ of $\msu(4)$ projected along $\L_2$ to the weight space of $\mso(5)$.}
 \label{fig:tetraedron}
\end{figure}

Some further remarks on the  group theory are in order.
The embedding $SO(5) \subset SO(6)$ is given  by extending the 
$SO(5)$ generators $M_{ij}, \ i,j\leq 5$  by $M_{i6}$.
This embedding is not regular,
i.e. the $SO(5)$ roots are not a subset of the $SO(6)$ roots.
However, the two commuting $SU(2)_L, SU(2)_R \subset SO(5) \subset SU(4)$ 
(correspond to the {\em long} roots of $SO(5)$) 
can be identified with the simple roots $\a_1, \a_3$ of $SU(4)\cong SO(6)$ which are orthogonal:
\begin{align}
 SU(2)_R \equiv SU(2)_{\a_1}, \qquad 
  SU(2)_L \equiv SU(2)_{\a_3} \ .
\end{align}
The  roots (and the weight lattice) of $SO(5)$ are then obtained by projecting 
the  roots (resp. the weights) of $SU(4)$ along $\L_2$ onto the $\a_1 \a_3$ - plane,
see figure \ref{fig:tetraedron}.
In particular, $SU(2)_R$ acts trivially on the highest weight (coherent) state 
$|\L\rangle$ of 
$\cH_N$, while $SU(2)_L = SU(2)_{\a_3}$ acts non-trivially and generates
 a $N+1$-dimensional representation 
at the edge of the weight tetrahedron of $\cH_N$. 
This gives precisely the fuzzy $S^2_{N+1}$ corresponding
to the fiber.
Hence at each point of $S^4$ there is an ``internal'' $SU(2)_L\subset H$ 
which  rotates the coherent states on $S^2$ over 
the same point of $S^4$.

\section{Some representation theory}

Representations of $\msu(4)$ are labeled by the Dynkin indices of their highest weights.
Weyls dimension formula gives
\begin{align}
 \dim (n,0,n) = \frac 1{12}(n+1)^2 (n+2)^2 (2n+3) \ .
\end{align}
For $\mso(5)$, let $\L = n\L_1 + n\L_2$ where $\a_1$ is the long root and $\a_2$ the short root.
The dimension of $(n,m)$ is given by 
\begin{align}
 \dim (n,m) = \frac 16 (n + 1)(m + 1)(n + m + 2)(2n + m + 3) 
\end{align}
or 
\begin{align}
 \dim (n-m,2m) = \frac 16 (2n + 3)(2m + 1)(n + m + 2)(n - m + 1) \ .
\end{align}
In particular,
\begin{align}
 \dim \cH_N = \dim (0,N) =  \frac 16 (N + 1)(N + 2)(N + 3) \ . 
 \label{dim-HN}
\end{align}
For example, 
\begin{align}
 \dim(1,0)&= 5, \quad 
 \dim(0,1)= 4, \quad
 \dim(0,2)=10 \ .
\end{align}
In particular $(0,2)$ is the adjoint.
One can decompose $(n,0,n)$ of $\msu(4)$ into irreps of $\mso(5)$.
It is not hard to see that 
\begin{align}
 (n,0,n) = (n,0) \oplus (n-1,2) \oplus (n-2,4) \oplus ... \oplus (0,2n) 
\end{align}
These are given by irreducible expressions of the type 
$P_n(X)\in (n,0), \ P_{n-1}(X)\cM \in (n-1,2), \ P_{n-2}(X)\cM\cM \in (n-2,4)$ and so on.

\end{document}